\documentclass[lettersize,journal]{IEEEtran}
\usepackage{cite}
\usepackage[caption=false,font=normalsize,labelfont=sf,textfont=sf]{subfig}
\usepackage{amsmath,amssymb,amsfonts}
\usepackage{graphicx}
\usepackage{textcomp}
\usepackage{xcolor}
\usepackage{booktabs}
\usepackage{multirow}
\usepackage{tabularx}
\usepackage{enumitem}

\usepackage{algorithm}
\usepackage{algorithmicx}
\usepackage{algpseudocode}

\begin{document}

\title{Federated Semi-Supervised and Semi-Asynchronous Learning for Anomaly Detection in IoT Networks\\}



\author{ Wenbin~Zhai, Liang~Liu, Feng~Wang, Youwei~Ding, Wanying~Lu, and Weizhi Meng\textsuperscript{†}
  \thanks{\textsuperscript{†}Corresponding author.}
    \thanks{ Wenbin~Zhai is with the Department of Computing, The Hong Kong Polytechnic University, Hong Kong. E-mail: wenbin.zhai@connect.polyu.hk}
  \thanks{Liang~Liu, Feng~Wang and Wanying~Lu are with the College of Computer Science and Technology at
    Nanjing University of Aeronautics and Astronautics, Nanjing, China. E-mails: \{liangliu, fengwang06, wanyinglu\}@nuaa.edu.cn}
  \thanks{Youwei~Ding is with the School of Artificial Intelligence and Information Technology at Nanjing University of Chinese Medicine, Nanjing, China.
    E-mail: ywding@njucm.edu.cn}
  \thanks{Weizhi~Meng is with the School of Computing and Communication at Lancaster University, UK. E-mail: weizhi.meng@ieee.org}
}

\markboth{Journal of \LaTeX\ Class Files,~Vol.~14, No.~8, August~2021}%
{Shell \MakeLowercase{\textit{et al.}}: A Sample Article Using IEEEtran.cls for IEEE Journals}

\IEEEpubid{0000--0000/00\$00.00~\copyright~2021 IEEE}

\maketitle

\begin{abstract}
  In recent years, Internet of Things (IoT) networks have become increasingly popular and widely used.
  However, they are vulnerable to malware, prompting effective techniques to detect infected IoT devices within the network.
  Federated Learning (FL) is a promising technique to balance privacy preservation, resource consumption and detection accuracy.
  Unfortunately, existing FL-based approaches are based on the unrealistic assumption that the data on the client-side is fully annotated with ground truths.
  Furthermore, it is a great challenge how to improve the training efficiency while ensuring the detection accuracy in the highly heterogeneous and resource-constrained IoT networks.
  Meanwhile, the communication cost between clients and the server is also a problem that can not be ignored.
  Therefore, in this paper, we propose a Federated Semi-Supervised and Semi-Asynchronous (FedS$^3$A) learning for anomaly detection in IoT networks.
  First, we consider a more realistic assumption that labeled data is only available at the server, and pseudo-labeling is utilized to implement federated semi-supervised learning,
  in which a dynamic weight of supervised learning is exploited to balance the supervised learning at the server and unsupervised learning at clients.
  Then, we propose a semi-asynchronous model update and staleness tolerant distribution scheme to achieve a trade-off between the round efficiency and detection accuracy.
  Meanwhile, the staleness of local models and the participation frequency of clients are considered to adjust their contributions to the global model.
  In addition, a group-based aggregation function is proposed to deal with the non-IID distribution of the data.
  Finally, the difference transmission based on the sparse matrix is adopted to reduce the communication cost.
  Extensive experimental results show that FedS$^3$A can achieve greater than 98\% accuracy even when the data is non-IID and is superior to the classic FL-based algorithms in terms of both detection performance and round efficiency, achieving a win-win situation.
  Meanwhile, FedS$^3$A successfully reduces the communication cost by higher than 50\%.
\end{abstract}


\section{Introduction}
\IEEEPARstart{O}{ver} the past few decades, with the development of sensors and wireless communication technologies,
Internet of Things (IoT) networks have become a popular architecture to support many modern applications or services,
such as smart homes, smart healthcare, smart cities, smart grid, and smart transportation, among many others \cite{MinFGCS}.
However, simultaneously with the massive deployment of IoT networks,
the attack surface where IoT devices may be hacked and exploited has grown in recent years,
such as DDoS attack, Side-channel attack or Bot attack \cite{IoTAttack1},
which seriously hinders the development of IoT networks \cite{IoTAttack}.
For this reason, effective countermeasures need to be developed to detect infected IoT devices in networks, in order to strengthen and ensure the security of IoT networks.

In recent years, artificial intelligence (AI) technologies,
such as machine learning (ML) and deep learning (DL), have been widely used for attack detection in IoT networks due to their superior performance \cite{AIIoT}.
Depending on the detection location, traditional AI-based schemes can be categorized into centralized and distributed detection \cite{IoTSSuvery}.
In centralized detection, IoT devices send their local data to a central server for detection.
This inevitably introduces expensive communication costs, serious privacy leakage, and poor scalability, although it can achieve high detection accuracy.
Distributed detection, known also as on-device detection, is performed locally on devices based on their private data.
Nonetheless, it is limited by the lack of interaction with and profit from other devices, which leads to the problem of isolated data islands, resulting in low detection accuracy \cite{COIoT}.

To address the aforementioned limitations, Federated Learning (FL) \cite{FedAvg} comes into being,
in which multiple devices (i.e., clients) cooperatively train a global model under the coordination of a central entity (i.e., the server) without the leakage of local data.
At each round of FL, clients perform the local training and upload the model parameters instead of raw data,
and then the server is responsible for aggregating and distributing the parameters to share the knowledge in the data of various clients,
thus achieving a trade-off between privacy, accuracy, communication cost and latency.
\IEEEpubidadjcol
The FL procedure continues until the model convergence or after a specified number of rounds.
Many FL-based approaches for anomaly detection in IoT networks have been proposed, and Table \ref{comparsion} summarizes some classic and state-of-the-art works \cite{DIoT,NBaIoTCN,ToNIoTCN,DADIoT,AFLIoT}.
Based on the comprehensive analysis, we find that the existing researches are still subject to the following challenges:

\begin{table*}[htbp]
  \caption{Comparison of Different Federated Learning Approaches}
  \begin{center}
    \scalebox{1.1}{
      \begin{tabular}{cccccc}
        \toprule
        \multirow{2}*{\textbf{ }}     & \textbf{Client}        & \textbf{Handling of}  & \textbf{Requirement for }     & \textbf{Communication} & \textbf{Communication Cost }   \\
                                      & \textbf{Participation} & \textbf{Non-IID Data} & \textbf{Training Data Labels} & \textbf{Mode}          & \textbf{Optimization}          \\
        \midrule
        Nguyen et al. \cite{DIoT}     & Full                   & No                    & Benign                        & Synchronous            & None                           \\
        Rey et al. \cite{NBaIoTCN}    & Full                   & Yes                   & Yes/Benign                    & Synchronous            & None                           \\
        Campos et al. \cite{ToNIoTCN} & Full                   & Yes                   & Yes                           & Synchronous            & None                           \\
        Liu et al. \cite{DADIoT}      & Full                   & No                    & Yes                           & Synchronous            & Gradient Compression           \\
        Tian et al. \cite{AFLIoT}     & One                    & Yes                   & Benign                        & Asynchronous           & Gradient Compensation          \\
        \textbf{Ours}                 & Partial                & Yes                   & Labels at Server              & Semi-Asynchronous      & Sparse Matrices of Differences \\
        \bottomrule
      \end{tabular}}
    \label{comparsion}
  \end{center}
\end{table*}

\begin{itemize}
  \item \textbf{Lack of Labeled Data:} A common limitation of existing works is that they only consider the supervised learning scenarios, where the local data is fully annotated with ground truths.
        It is not realistic because users and end devices do not have the ability and corresponding expertise to label the data \cite{FedMatch}.
        Furthermore, some works \cite{DIoT,AFLIoT} argue their schemes are unsupervised, but in fact, they are based on the assumption that all data in the model training phase is benign, which is essentially supervised.
        At the same time, this assumption is unreasonable because we can not guarantee it in real networks,
        and once the devices are compromised during the training phase, the malicious samples will seriously degrade the performance \cite{Backdoor}.
  \item \textbf{Device Heterogeneity:} Although synchronous FL \cite{DIoT,IoTSuvery1,IoTFL1} is a natural choice, due to the vast differences in computing capacities, communication resources and data volumes,
        the required time for different devices to perform local updates and receive/upload models varies greatly \cite{SAFA}.
        Therefore, due to the device heterogeneity, the server has to wait for the slowest client in synchronous FL, which leads to unnecessary waiting time and low round efficiency.
        Although asynchronous FL can deal with the above problems, it requires frequent model updates, resulting in an expensive consumption of communication and computing resources.
        The server receives numerous local updates sent by clients, which can overwhelm the server and provide little benefit to the model convergence \cite{FedSA}.
  \item \textbf{Non-IID Data:} In IoT networks, some devices may function normally, while others may suffer from multiple attacks.
        As a result, the local data of devices can not be regarded as a uniform distribution of the overall sample.
        In other words, the data of different clients is often not independent-and-identically-distributed (Non-IID) \cite{NonIID2}.
        It has been proved that in the presence of non-IID data, the performance of FL degrades significantly in terms of the convergence rate and model accuracy \cite{NonIID1}.
  \item \textbf{Significant Communication Cost:} Although FL does not need to upload local raw data to the server,
        frequent updates of model parameters (up to tens or even hundreds of rounds) are still a huge communication overhead for resource-constrained IoT networks.
        Meanwhile, the server may become a communication performance bottleneck since it needs to frequently distribute/receive the models to/from clients \cite{CEFL1}.
\end{itemize}

To overcome the deficiencies described above, in this paper,
we propose a Federated Semi-Supervised and Semi-Asynchronous (FedS$^3$A) learning for anomaly detection in IoT networks.
In particular, the main contributions of this paper are summarized as follows:
\begin{itemize}
  \item We consider a more realistic assumption that each client has massive local unlabeled data and the server stores a little labeled data,
        which is different from the previous scenarios where the client data is all attached with ground-truths while there is no data at the server.
        Meanwhile, we propose a novel federated semi-supervised learning system for anomaly detection in IoT networks.
        Specifically, the pseudo-labeling technique is used to label the data at clients, and FL is combined for model training, in which clients perform unsupervised training and the server implements supervised learning.
  \item We propose a semi-asynchronous federated learning scheme, consisting of the semi-asynchronous model update and staleness tolerant distribution,
        due to the heterogeneity of IoT devices, to achieve a good trade-off between client utilization, round efficiency and model performance.
        Unlike synchronous FL, FedS$^3$A implements the global update once the server receives the $C$ proportion of local models instead of all local models.
        Meanwhile, FedS$^3$A allows the local models of a subset of clients to remain asynchronous with that of the server, in order to take full advantage of the data and training progress of each client.
  \item We design a novel aggregation function, since FedS$^3$A considers a new federated semi-supervised and semi-asynchronous learning scenario in this paper. 
        First, a dynamic weight of supervised learning is employed to balance its guiding role for unsupervised learning and adverse impact on model overfitting.
        Then, the weights of stale models are reduced to alleviate the poisoning of the global model.
        Meanwhile, a group-based aggregation function is proposed to minimize the impact of non-IID data on the global model, which can capture the main gradient features under different data distributions.
        Furthermore, We adopt an adaptive learning rate based on the participation rounds of clients to balance their contributions to the global model.
  \item We design a simple but effective approach to reduce the communication cost without the sacrifice of the model performance,
        by sending only the difference of parameters as the sparse matrices across the communication rounds.
  \item We implement extensive experiments on the CIC-IDS 2017 \cite{CIC-IDS} dataset.
        The experimental results show that the performance of FedS$^3$A is far superior to the state-of-the-art detection methods \cite{DIoT,NBaIoTCN,ToNIoTCN,DADIoT,AFLIoT} based on federated learning \cite{FedAvg,AFL} in terms of the model performance, communication cost and round efficiency.
        It can achieve greater than 98\% accuracy even when the data is non-IID and reduce higher than 50\% communication cost.
        Furthermore, we conduct ablation experiments on FedS$^3$A and evaluate the impact of different hyperparameters and functions to demonstrate its efficiency and effectiveness.
\end{itemize}

The remainder of the paper is organized as follows.
In Section \ref{RelatedWork}, we review and summarize the existing works whose topics are related to this paper.
Section \ref{Model} introduces the federated learning procedure and formalizes the system model.
The proposed FedS$^3$A is described in Section \ref{FedS$^3$A} in detail.
In Section \ref{Experiment}, the performance of FedS$^3$A is evaluated through extensive experiments.
Finally, Section \ref{Conclusion} concludes this paper.

\section{Related Work}\label{RelatedWork}
In this section, we review related works in literature.
First, the applications of FL for anomaly detection in IoT networks are introduced.
Then, we summarize the optimizations of FL from the perspectives of data labeling and model update,
namely federated semi-supervised learning and semi-asynchronous federated learning.

\subsection{Federated Learning for Anomaly Detection in IoT Networks}
Federated Learning has been widely used for anomaly detection in IoT networks due to cooperative learning and privacy preservation.
DIoT \cite{DIoT} is the first attempt to apply the federated learning approach for anomaly detection in IoT networks.
It assumes that IoT devices are not compromised during the model training phase, leaving enough time to utilize the Gate Recurrent Unit (GRU) model to learn benign behavior patterns.
Abnormalities are detected by deviations from benign behaviors.
Liu \textit{et al.} \cite{DADIoT} use Convolutional Neural Networks (CNN) to extract fine-grained features, and then the trained Long Short-Term Memory (LSTM) model is utilized for anomaly detection.
In addition, in order to improve communication efficiency, a top-k selection-based gradient compression scheme is proposed to reduce the communication cost.

Rey \textit{et al.} \cite{NBaIoTCN} studies the performance of federated learning in two scenarios based on the N-BaIoT \cite{N-BaIoT} dataset.
One is that the client-side data is all labeled, and MultiLayer Perceptron (MLP) is used to perform supervised learning for classification,
while the other is that clients only have benign data and utilize AutoEncoder (AE) for anomaly detection training.
Based on the CIC-ToN-IoT \cite{CIC-ToN-IoT} dataset, Campos \textit{et al.} \cite{ToNIoTCN} define three scenarios, namely basic, balanced, and mixed scenarios, according to the distribution of the client-side dataset.
Meanwhile, a novel aggregation function called Fed+ is proposed, which is more suitable for non-IID and skewed training data, and Multiple Linear Regression (MLR) is used for model training.
Tian \textit{et al.} \cite{AFLIoT} focus on the gradient divergence and delay problem caused by the IoT device heterogeneity and non-IID data in federated learning.
A method based on Taylor expansion is proposed to compensate for the inconsistency.

\subsection{Federated Semi-Supervised Learning}
Although researchers have made great efforts in federated learning for anomaly detection,
existing frameworks make an unrealistic assumption that client-side data comes with ground-truth labels.
In fact, the data of clients is often unlabeled in IoT networks, which may be due to the lack of expertise, the consideration of privacy protection and the high cost of labeling.

Semi-supervised learning \cite{Fixmatch} has attracted increasing attention in the past few years.
It aims to learn a model under massive unlabeled data with a little supervised information.
Therefore, in the context of federated learning, a novel and promising paradigm called federated semi-supervised learning (FSSL) is proposed and quickly becomes a research hotspot \cite{lu2023framework}.
There are two essential scenarios according to the location of the labeled data.
The first is the conventional scenario where clients have both labeled and unlabeled data (labels-at-client),
and the second is the disjoint scenario where the labeled data is only available at the server (labels-at-server) \cite{FedMatch}.

Zhu \textit{et al.} \cite{GPSFSSL} combine the CNN-GRU model with the pseudo-labeling technique for FSSL to identify the travel model.
Meanwhile, a group-based aggregation function and a flip-based data augmentation (DA) scheme are proposed to improve the performance and convergence speed of the trained model.
Nandury \textit{et al.} \cite{DLFSSL} analyze the influence of the update diversity of model parameters on model convergence,
and propose diversity scaling to compensate for the effect of inter-client update dissimilarity.
Zhang \textit{et al.} \cite{Benchmarking} propose a metric to measure the non-IID distribution of inter-client data.
Pseudo-labels generated by weak DA samples are used to train strong DA samples.
Meanwhile, a group-based model aggregation is proposed to achieve a trade-off between supervised and unsupervised learning.

\subsection{Semi-Asynchronous Federated Learning}
Synchronous federated learning has become a natural choice, however, due to the device heterogeneity,
the server has to wait for the slowest client in the synchronous protocol, which leads to low round efficiency, slow convergence speed, and low client utilization.
To alleviate the above issues, asynchronous federated learning is proposed,
whose essential idea is to perform aggregation once the server receives the model parameter uploaded by one client, without waiting for other clients,
while allowing clients to use the stale global model for local training \cite{AFL}.
Stale control has been shown to be the key to guaranteeing model convergence.

However, asynchronous federated learning requires frequent global aggregations and model updates,
which consumes a lot of communication and computing resources with little benefit for model convergence.
Furthermore, the rapid staleness of client-side models will lead to the reduction of training accuracy, especially on non-IID data.
Therefore, semi-asynchronous federated learning comes into being, which relaxes the requirement on the number of participating clients at each round,
that is, the global aggregation can start when the server receives the parameters from a part of the clients.
It is obvious that asynchronous and synchronous federated learning can be regarded as special cases of semi-asynchronous federated learning.

Wu \textit{et al.} \cite{SAFA} allow some lag tolerance to leverage the knowledge of clients,
while forcing to synchronize seriously outdated clients to prevent them from poisoning the global model.
Ma \textit{et al.} \cite{FedSA} conduct a theoretical analysis of the quantitative relationship between the convergence boundary of federated learning and different factors,
such as the number of participating clients in each round, the degree of data non-IID distribution, and the heterogeneity of clients.
Meanwhile, an adaptive learning rate based on client participation is designed to improve training accuracy.
Chen \textit{et al.} \cite{AFL1} propose a temporally weighted aggregation function, i.e., the most recently updated model should have a higher weight in the aggregation.
In addition, the parameters of each layer of the CNN model are further analyzed to reduce the communication overhead.
Chai \textit{et al.} \cite{FedAT} stratify the clients according to the training speed, using the synchronous aggregation within the layer and asynchronous update between layers.
Meanwhile, a lossy compression technique based on polyline encoding is adopted to reduce the communication cost.


\section{Preliminaries and System Model}\label{Model}
In this section, we first give a systematic introduction to federated learning, and then the system model is described.

\subsection{Federated Learning}\label{FL}
Federated Learning is first proposed in \cite{FedAvg}, whose idea is to allow users to cooperatively train a global model in order to profit from data of peer entities without sharing local raw data, i.e., ensuring data local privatization.
A traditional federated learning system consists of a centralized server $S$ and multiple decentralized clients (ranging from several to thousands) $C_i, 1 \leq i \leq M$.
Clients perform local training based on their private data, and cooperatively train a global model under the coordination and command of the server, which acts as an aggregator.
The procedure of federated learning is outlined as follows:

\subsubsection{Model Initialization}
At the round $r$ of the training, the server $S$ randomly selects partial or all of the clients $C_i, 1\leq i \leq N \leq M$ to participate in this round of the training,
and then sends the global model parameters $\omega_g^r$ to these designated clients.
At the first round, the global model parameter is specified or initialized randomly, otherwise it is aggregated from the local model parameters of clients.

\subsubsection{Local Training}
Each client participating in this round of training $C_i, 1\leq i \leq N \leq M$ performs $E$ epochs of supervised-learning training on the basis of its local training data sample $D_i = \{(x_j, y_j) \mid 1 \leq j \leq |D_i| \}$,
where $x_j$ is the feature vector and $y_j$ is the corresponding one-hot label.
The training objective of the client is to minimize the following objective function:
\begin{equation}
  \mathop{\arg\min}\limits_{\omega_i^r} \ F_i(\omega_i^r) = \frac{1}{|D_i|} \sum_{j=1}^{|D_i|} f(\omega_i^r, x_j, y_j)
\end{equation}
where $\omega_i^r$ is the model parameter of client $C_i$ at round $r$,
and $f(\omega_i^r, x_j, y_j)$ is the loss function, which is used to measure the loss deviation between the prediction and the ground-truth of the data sample under the current model parameter $\omega_i^r$,
such as cross-entropy loss \cite{CEL}.

Many optimization methods such as stochastic gradient descent (SGD) \cite{SGD} or Adam \cite{Adam} are used to find the optimal solution.
The model parameter $\omega_i^r$ is updated locally by
\begin{equation}
  \omega_i^{r+1} = \omega_i^r - \eta_i^r \nabla F_i(\omega_i^r)
\end{equation}
where $\eta_i^r$ is the learning rate of $C_i$ at round $r$, and $\nabla F_i(\omega_i^r)$ is the gradient of the loss function with respect to $\omega_i^r$.

After completing the local training, each participating client $C_i$ uploads the updated model parameter $\omega_i^{r+1}$ to the server $S$.

\subsubsection{Global Aggregation}
The server $S$ performs the aggregation operation using a specific aggregation function, such as FedAvg \cite{FedAvg}, after receiving the local updated parameters of all selected clients to obtain the global model parameter for the next round of training.
The aggregation function of FedAvg can be represented as
\begin{equation}
  \omega_g^{r+1} = \sum_{i=1}^{N} \frac{|D_i|}{|D|} \omega_i^{r+1}
\end{equation}
where $D = \sum_{i=1}^{N} D_i$ and $D_i \cap D_{i'} = \emptyset$, for any $i \neq i'$.
Meanwhile, the global training objective can be defined as
\begin{equation}
  \mathop{\arg\min}\limits_{ \omega_g^{r+1}} \ F_g(\omega_g^{r+1}) = \frac{\sum_{i=1}^N D_i F_i(\omega_g^{r+1})}{D}
\end{equation}

Repeat the above process until the global model converges or the training round reaches the specified maximum number of rounds $R$.

\begin{figure*}[htbp]
  \centering
  \includegraphics[width=0.8\textwidth]{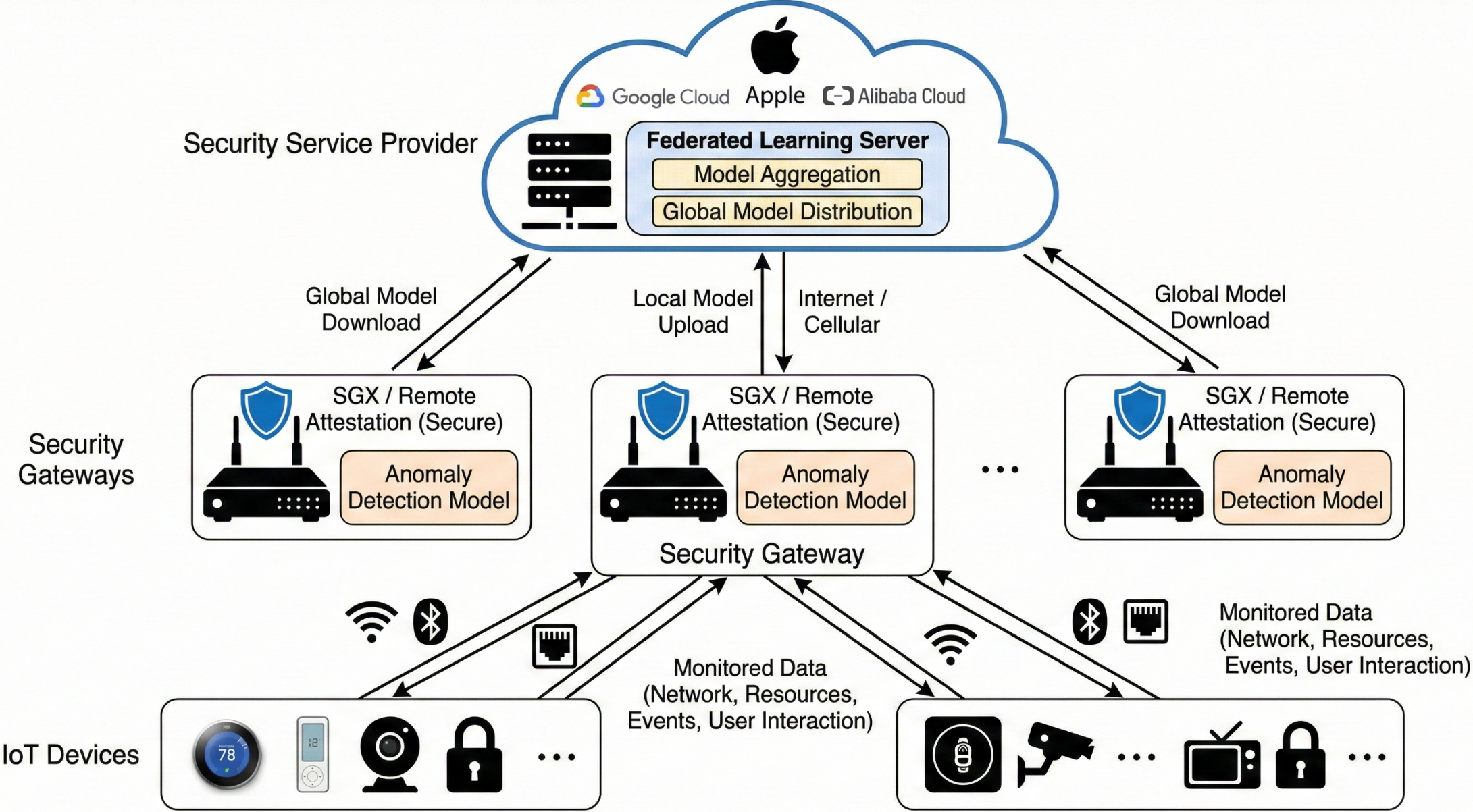}
  \caption{The system architecture of the IoT network.}
  \label{Fig.Architecture}
\end{figure*}

\subsection{System Architecture}
The IoT network presented in this paper consists of IoT devices, security gateways, and a security service provider.
The system architecture is depicted in Fig. \ref{Fig.Architecture}.

\subsubsection{IoT Devices}
IoT devices are connected to the corresponding security gateway through Wi-Fi, Bluetooth or Ethernet.
They are vulnerable and may be attacked, compromised or exploited by adversaries.
Therefore, various behaviors of IoT devices, such as network communication, resource consumption, software event, and user interaction,
are monitored and collected by the security gateway for further anomaly analysis and detection.

\subsubsection{Security Gateways}
Since IoT devices are moderately reliable and have limited resources,
in this paper, the entity deploying and holding the anomaly detection component is not the IoT devices that need to be protected,
but more powerful and reliable security gateways.
Security gateways have sufficient resources to use effective approaches, such as Intel SGX \cite{SGX} or remote attestation techniques \cite{RemoteAttestation}, to ensure that they are not compromised.
Security gateways act as the local access gateways, monitor the communication, network traffic or other data of the connected IoT devices,
and use the trained model for anomaly detection.
Meanwhile, they connect to the security service provider via the Internet or cellular,
as clients in the federated learning system,
uploading and downloading the models under the coordination of the security server provider.

\subsubsection{Security Service Provider}
The security service provider, i.e., the server of the FL system, is often held by a powerful organization, such as Google, Apple, Alibaba, and it is responsible for the overall coordination and command of anomaly detection,
such as the aggregation of local models and distribution of the global model.

\section{FedS$^3$A Mechanism}\label{FedS$^3$A}
In this section, we introduce our proposed Federated Semi-Supervised and Semi-Asynchronous (FedS$^3$A) learning mechanism for anomaly detection in IoT networks.
\subsection{Core Workflow}
\begin{figure*}[htbp]
  \centering
  \includegraphics[width=0.8\textwidth]{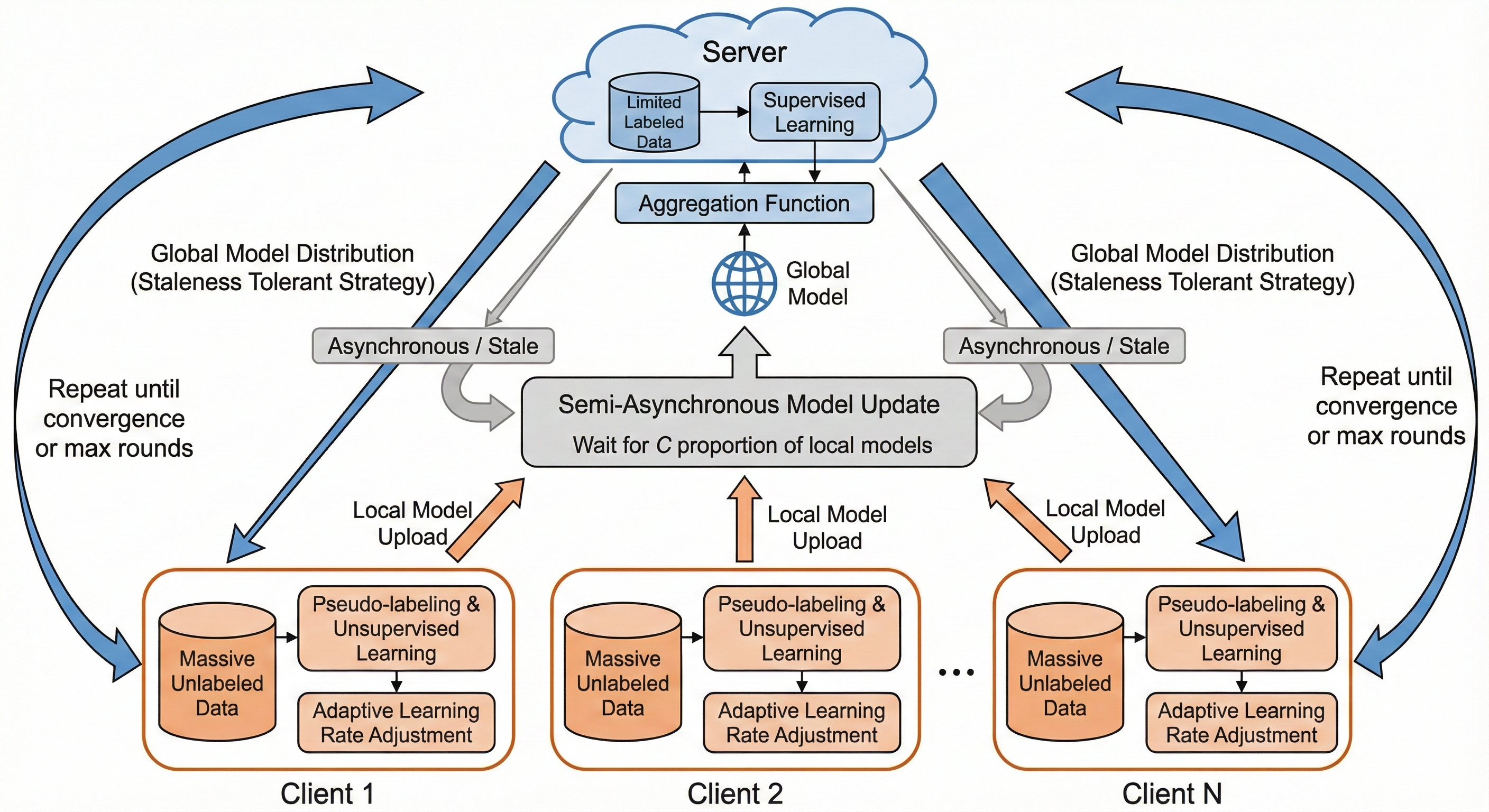}
  \caption{The core workflow of FedS$^3$A.}
  \label{Fig.Workflow}
\end{figure*}

Fig. \ref{Fig.Workflow} shows the main workflow of FedS$^3$A.
First, since we consider a more realistic assumption in this paper, namely all data at clients is unlabeled and the server has a little labeled data,
we propose to utilize federated semi-supervised learning for model training, in which clients perform unsupervised learning through the pseudo-labeling technique based on the massive local unlabeled data,
while the server utilizes very limited labeled data for supervised learning, which will be introduced in Section \ref{FSSL}.
Then, in order to achieve a good trade-off between client utilization, round efficiency and model performance,
we design a semi-asynchronous model update scheme, that is, all clients are allowed to participate in the model training, and the global update will be implemented once the server receives the $C$ proportion of local models instead of all local models.
The details can refer to Section \ref{SAMU}.
Meanwhile, the design of the model aggregation function is described in Section \ref{AF}.

Next, a staleness tolerant distribution strategy is used for the model distribution.
Specifically, we allow the local models of a subset of clients to remain asynchronous with that of the server, 
in order to take full advantage of the data and training progress of each client.
In other words, we only implement model distribution and update for some clients, which will be introduced in detail in Section \ref{STD}.
Furthermore, we also adaptively adjust the learning rate of the client according to the participation rounds in global updates in order to balance the contributions of different clients to the global model,
whose details will be described in Section \ref{adr}.
The above process will be repeated until the global model converges or the training round reaches the specified maximum number of rounds.

\subsection{Federated Semi-Supervised Learning}\label{FSSL}
Due to the superior performance and privacy preservation of federated learning,
it has been widely used for anomaly detection in IoT networks.
However, as shown in Table \ref{comparsion}, most of the existing advanced schemes are based on the traditional federated learning system,
in which clients perform supervised-learning training, which means the data stored at each client is fully annotated with ground-truth labels.
Meanwhile, the server has no data and acts only as an aggregator and coordinator.
In fact, this is not realistic in real-world applications for IoT networks.
Massive amounts of data are constantly generated in IoT networks, and users have no expertise to annotate data while the server has the corresponding capacity to obtain a certain amount of labeled data.

In addition, the other works are based on the assumption that ``IoT devices are not compromised in the model training stage, leaving sufficient time to learn models of benign behaviors".
Unfortunately, this assumption is not reasonable for vulnerable IoT devices with limited energy and computing resources.
Once the devices are attacked and compromised during the training phase,
malicious training data which is mistaken as benign will greatly reduce the accuracy of the model,
whose effect is similar to data poisoning (label-flipping) attacks \cite{DataPoisoning}.

Motivated by this, in this paper, we make a more realistic assumption and consider a new federated learning scenario for IoT networks, called the disjoint scenario for federated semi-supervised learning (FSSL),
where labeled data is only available at the server, and each client has a large amount of unlabeled data.
The details of FSSL and differences from traditional federated learning are described as follows.

\subsubsection{Unsupervised Learning at Clients}
Due to the lack of professional knowledge and the huge scale of private data, client-side data samples are usually not attached with any labels.
Therefore, in this paper, we adopt a pseudo-labeling-based \cite{Pseudo} training method to achieve unsupervised training on the unlabeled dataset at the clients.
The loss function is defined as follows:
\begin{equation}
  F_i(\omega_i^r) = \frac{1}{|D_i|} \sum_{j=1}^{|D_i|} \operatorname{sgn} (\max(\bar{y}) \geq \theta) \ l(\arg\max (\bar{y}), \bar{y})
\end{equation}
where $ \bar{y} =  p(\omega_i^r, x_j)$, which means the prediction of the current model parameter $\omega_i^r$ on the sample $x_j$, $l( \ )$ is the cross-entropy loss function,
$\operatorname{sgn} ( \ )$ is the indicator function, and $\theta$ is the threshold hyperparameter,
which is used to determine which samples with high confidence can be attached with pseudo-labels and used to help train the model.

Recall that, on the client side, the data is all unlabeled.
Therefore, using this approach, we can utilize the generated pseudo-labels to help train the model,
which can improve the performance of the model due to the utilization of large and diverse data on the client side.

\subsubsection{Supervised Learning at the Server}
It is well known that it is difficult to train a good model solely on unlabeled data of clients, especially on non-IID data.
Fortunately, in real-world applications for IoT networks, the security service provider (i.e., the server) is usually hosted by an organization, such as Google, Apple, Alibaba, which has the capacity to obtain a certain amount of label data samples.
Therefore, at each round of training, after distributing the global model parameter to each client, the server also use this parameter for supervised learning based on the local labeled data.
The loss function can be represented as
\begin{equation}
  F_s(\omega_s^r) = \frac{1}{|D_s|} \sum_{j=1}^{|D_s|} l(y_j, p(\omega_s^r, x_j))
\end{equation}
where $\omega_s^r$ is the model parameter of the server $S$ at round $r$ and $D_s$ is the labeled dataset of the server.

The disjoint scenario for FSSL we considered in this paper is more in line with real IoT network scenarios.
At the same time, through the pseudo-labeling technique, our method better combines the client-side unsupervised learning and server-side supervised learning for federated-learning training.
Specifically, under the guidance of a small amount of labeled data on the server,
our method makes full use of the massive unlabeled data on the clients,
thereby improving the learning performance of the model and avoiding the waste of data resources.
It efficiently solves the problems that the generalization of supervised learning is poor when there is very limited labeled data,
and the accuracy of unsupervised learning is low when there is no labeled data guidance.

\subsection{Semi-Asynchronous Federated Learning}\label{SAFL}
Although synchronous FL has been widely used for anomaly detection in IoT networks,
several limitations stand out as follows:
1) Low client utilization. Clients participating in each round of training are pre-specified and randomly selected.
As a result, many capable clients remain idle even when they are ready and willing to participate in model training \cite{SAFA}.
2) Low round efficiency. Before aggregation at each round, the server has to wait for all selected clients to finish local training or reach the timeout threshold,
however, due to the high heterogeneity of IoT devices, the time for devices to perform local updates varies greatly, resulting in low round efficiency.
3) Waste of training progress. Selected clients may not complete local training in time, and their progress may be wasted
as clients participating in the next round of training are reassigned and their local models are forced to be overwritten by the global model.

Asynchronous FL is proposed to address the above shortcomings, but it also introduces the following new problems:
1) Significant computational and communication cost: In asynchronous FL, the server starts the global update upon receiving a model parameter from an arbitrary client,
which results in significant resource consumption.
2) Bias in client contributions: due to the high heterogeneity of devices, clients with faster computing will frequently participate in the global update,
resulting in a greater impact on the trained global model.
3) Poisoning of staleness: Some local models with large staleness can poison the global model, resulting in lower training accuracy, especially on non-IID data \cite{StaleNon-IID}.

Therefore, in order to achieve a trade-off between the client utilization, model accuracy, communication cost, round efficiency and convergence rate,
in this paper, based on the FSSL, we further design a semi-asynchronous federated learning scheme,
which consists of the semi-asynchronous model update and staleness tolerant distribution.

\subsubsection{Semi-Asynchronous Model Update}\label{SAMU}
First, we design a semi-asynchronous model update approach that utilizes a hyperparameter $C$ to control the proportion of clients participating in each round of the global aggregation.
This idea has been widely applied to anomaly detection in IoT networks, for instance, the most existing methods \cite{DIoT,NBaIoTCN,ToNIoTCN,DADIoT} use FedAvg \cite{FedAvg} for the global model update,
whose idea is to select the $C$ proportion of clients in advance to participate in each round of training, and then wait for all selected clients to complete the training before the global aggregation,
which is essentially a synchronous method.

In our approach, we retain the hyperparameter, however, unlike the previous methods, we no longer pre-specify participating clients, but instead allow all clients to participate if they are willing to.
Then, the server starts the model aggregation once the $C$  proportion of local updates from clients have been received.
It is obvious that the efficiency of FL is closely related to the participation proportion of clients.
Meanwhile, synchronous and asynchronous FL can be regarded as special cases of semi-asynchronous FL (i.e., when $C$ is close to 1 and 0).

Due to the device heterogeneity, some devices may not be able to complete local updates in time (i.e, drop out the $C$ proportion) due to the large data size or poor computing capacity.
However, in fact, their data is valuable and should be fully utilized in the federated learning procedure to improve the model quality.
Therefore, we further propose a staleness tolerant distribution method.

\subsubsection{Staleness Tolerant Distribution}\label{STD}
In federated learning, version control is by no means easy.
The synchronous FL distributes the latest model at each round, which simplifies the FL process, but also inhibits the possibility of accelerating the convergence.
Furthermore, if the above semi-asynchronous model update is used solely,
some local models will never be able to participate in the procedure of FL, reducing the accuracy of the trained model, especially in highly heterogeneous IoT networks.
Therefore, in order to fully utilize the data of different clients, in this section,
we introduce a staleness tolerant distribution method to accompany the semi-asynchronous model updates.

Our approach does not enforce synchronization between clients and the server.
In other words, we allow some clients to remain asynchronous with the server, and tolerate outdated local models (i.e., allow the version of some local models to be inconsistent with that of the server).
The idea behind is to encourage clients with the outdated model to participate in the aggregation, and use their data and progress to make the training of federated learning faster and better.
In the following, we detail our staleness tolerant distribution strategy and show how it works with the semi-asynchronous model update.

According to the gap in the version of each client and the server, and whether the client participate in the global aggregation, we can divide the clients into three types, namely \emph{latest}, \emph{tolerable} and \emph{deprecated} clients, which are defined as follows:
\begin{enumerate}
  \item \emph{Latest Clients:} The clients that complete local training and successfully participate in the semi-asynchronous global model update at this round.
  \item \emph{Deprecated Clients:} The clients whose local model version is too stale compared to the latest global model, possibly because of the network failure or system crash.
  \item \emph{Tolerable Clients:} The clients that do not perform local training based on the latest global model, but the model version they are based on is not too stale either and acceptable.
        This is a state between \emph{Latest} and \emph{Deprecated}.
\end{enumerate}

As mentioned earlier, at the round $r$, the server implements the model aggregation once the $C$ proportion of local updates from clients have been received, and then the global model parameters of the next round $\omega_g^{r+1}$ can be obtained.
Let $\omega_i^{r_i}$ denote the model parameter based on which client $C_i$ performs the local training, i.e., the version of the local model of $C_i$ is $r_i$.
In synchronous FL, $r_i = r$, for any client $C_i, 1 \leq i \leq M$, while $r_i \leq r$ in semi-asynchronous and asynchronous FL.
The server then implements a staleness tolerant distribution strategy that performs different actions on different types of clients based on the model version differences and global update participation, which can be described as follows.

First, for clients that participated in the semi-asynchronous model update at the current round (i.e., \emph{latest} clients),
we distribute the latest global model parameters $\omega_g^{r+1}$ to them regardless of their local model version.
Then, for each other client $C_i$, we judge whether the version gap between its local model $\omega_i^{r_i+1}$ and the global model $\omega_g^{r+1}$ is greater than a threshold $\tau$.
If so (i.e., $r-r_i > \tau$), it is a \emph{deprecated} client and need to be forced to update, so the server distributes the latest global model to it.
Otherwise, it is a \emph{tolerable} client (i.e., $r-r_i \leq \tau$), and the server allows it to remain asynchronous and does not force the distribution of the latest global model.

With the combination of the semi-asynchronous model update and staleness tolerant distribution, we achieve the version control of the local models.
We synchronize the latest clients with the server to prevent model divergence.
At the same time, deprecated clients are forced to update, so that the global model will not be poisoned by the heavily outdated local models.
Furthermore, we allow tolerant clients to remain asynchronous with the server, so that the data and knowledge of each client can be fully utilized to improve the performance.
Model staleness control has been shown to be the key to guaranteeing convergence \cite{FedAT}.
Poorly controlled staleness, such as being too small, can cause some slower clients to have large staleness before completing their local training.
They will be easily forced to synchronize.
Therefore, these clients may never participate in global updates, which reduces the training accuracy, especially on non-IID data.

\begin{figure*}[htbp]
  \centering
  \includegraphics[width=0.95\textwidth]{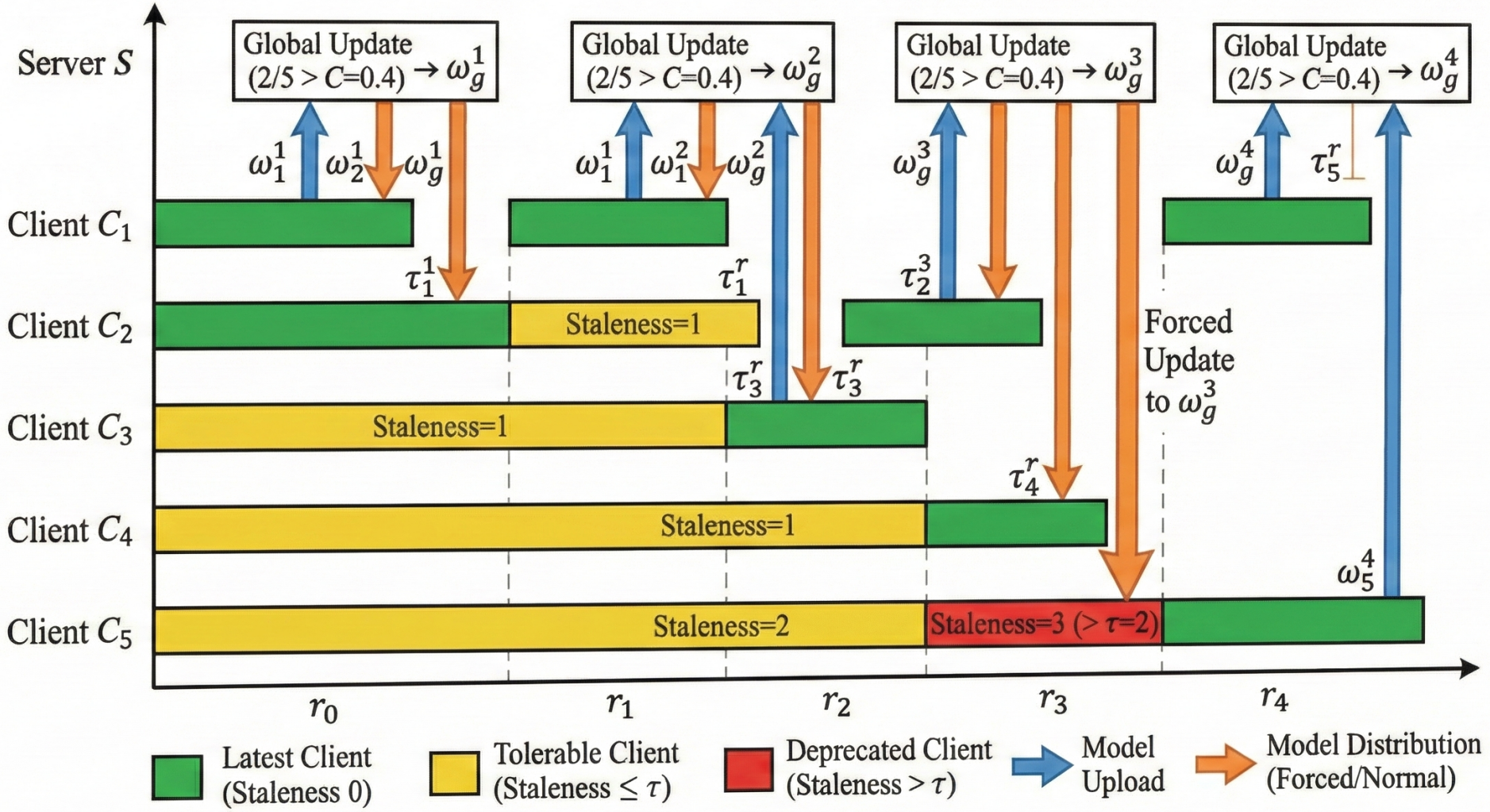}
  \caption{Illustration of the FedS$^3$A mechanism when $C  = 0.4$ and $\tau = 2$.}
  \label{Fig.Asyn}
\end{figure*}

For instance, as shown in Fig. \ref{Fig.Asyn}, there are five clients $C_1 \sim C_5$ and a server $S$ with $C = 0.4, \tau = 2$.
Clients perform unsupervised learning, while the server performs supervised learning.
At the round $r_0$, clients $C_1$ and $C_2$ are the first to complete local training and upload model parameters $\omega_1^1$ and $\omega_2^1$ to the server $S$.
Since the server has received the parameters from $C*M = 0.4 * 5 =2$ clients,
the new global model $\omega_g^1$ is updated by these two unsupervised parameters and the supervised parameter of the server $\omega_s^1$.
Meanwhile, the staleness of each client at different rounds is shown in Table \ref{stale}.
The staleness of $C_1$ and $C_2$ is $0$ due to the participation at this round while the staleness of others is $1 < \tau = 2$, which means they are tolerable clients and do not need to be forced to update.
Then, at the round $r_1$, $C_1$ and $C_3$ participate in the global update, so their staleness is updated to $0$, while the staleness of other clients is increased by one.
Next, at the round $r_2$, $C_2$ and $C_4$ successfully complete the local training.
Meanwhile, it is worth noting that the staleness of  $C_5$ becomes $3 > 2$, which may be due to the network failure or system crash.
Therefore, $C_5$ is a deprecated client and the server needs to distribute the latest global model $\omega_g^3$ to it.
Finally, $C_5$ successfully participates in the global update at the round $r_3$.

\begin{table}[htbp]
  \caption{Staleness of clients at different rounds}
  \begin{center}
    \scalebox{1.2}{
      \begin{tabular}{cccccc}
        \toprule
        \textbf{ } & \textbf{$C_1$} & \textbf{$C_2$} & \textbf{$C_3$} & \textbf{$C_4$} & \textbf{$C_5$} \\
        \midrule
        $r_1$      & $\tau_1^1 = 0$ & $\tau_2^1 = 0$ & $\tau_3^1 = 1$ & $\tau_4^1 = 1$ & $\tau_5^1 = 1$ \\
        $r_2$      & $\tau_1^2 = 0$ & $\tau_2^2 = 1$ & $\tau_3^2 = 0$ & $\tau_4^2 = 2$ & $\tau_5^2 = 2$ \\
        $r_3$      & $\tau_1^3 = 1$ & $\tau_2^3 = 0$ & $\tau_3^3 = 1$ & $\tau_4^3 = 0$ & $\tau_5^3 = 3$ \\
        $r_4$      & $\tau_1^4 = 2$ & $\tau_2^4 = 1$ & $\tau_3^4 = 0$ & $\tau_4^4 = 1$ & $\tau_5^4 = 0$ \\
        \bottomrule
      \end{tabular}}
    \label{stale}
  \end{center}
\end{table}

\subsection{Design of Aggregation Function}\label{AF}
Most of the existing FL-based anomaly detection schemes for IoT networks use FedAvg as the aggregation function, which is the most dominant method and has been proven to be convergent \cite{FedAvgConPro}.
In this paper, since we consider a novel semi-supervised learning and semi-asynchronous model update scenario,
which is different from that of the existing works, a new aggregation function should be proposed.
A naive approach is to simply transfer FedAvg over, as shown below.
\begin{equation}
  \omega_g^{r+1} = \frac{|D_s|}{|D_c| + |D_s|} \omega_s^{r+1} + \sum_{i=1}^{C*M} \frac{|D_i|}{|D_c| + |D_s|} \omega_i^{r_i+1}
\end{equation}
where $|D_c| = \sum_{i=1}^{C*M} |D_i|$.
It retains the weighted average idea of FedAvg, while taking the supervised-learning training at the server into consideration.
However, due to the particularity of the scenario, there are still some challenges that need to be carefully considered.

\subsubsection{The weight of supervised learning needs to be revisited}
As mentioned earlier, in the FSSL-based scenario for anomaly detection in real-world IoT networks,
a large amount of data is unlabeled, while labeled data is very scarce and usually only available at the server.
However, labeled data is valuable, and its impact on the model performance is undoubtedly significant.
Therefore, in this paper,
we use a dynamic supervised learning weight and fit it through a function $f(r)$, which satisfies the following conditions:
\begin{enumerate}
  \item $0 < f(r) < 1, \forall \ 0< r \leq R $.
  \item In the early stage of the training, $f(r)$ should take a large value $\alpha$, such as $\alpha = \frac{1}{2}$.
        The reason is that at the beginning, the model and clients have little knowledge of the data,
        and the supervised-learning training at the server has an important guiding significance for the unsupervised learning at clients.
  \item As the number of rounds increase, $f(r)$ decreases monotonically and gradually approaches a constant $\beta$ (i.e., $\lim _{r \rightarrow R} f(r) = \beta$).
        This is because with the increase of the number of rounds and improvement of the model performance,
        the weight of supervised learning should be reduced to prevent the model from overfitting.
        At the same time, a large amount of unlabeled data on the client side can be fully utilized in this way to improve the learning performance and avoid the waste of data resources.
  \item The value of $\beta$ is significant. If it is set too large, it will lead to overfitting; if it is too small, it will cause the model to forget the knowledge learned from the labeled data.
        In this paper, we set $\beta$ to $\frac{1}{C*M + 1}$, namely we equate the weight of the server to the average weight of clients in the end.
\end{enumerate}


Based on the reconsideration of the supervised learning weight, the aggregation function can be rewritten as

\begin{equation}
  \omega_g^{r+1} = f(r) * \omega_s^{r+1} + (1-f(r)) * \sum_{i=1}^{C*M} \frac{|D_i|}{|D_c|} \omega_i^{r_i+1}
\end{equation}

\subsubsection{The weights of stale models need to be reduced}\label{staleness}
The above aggregation function measures the importance of each local model based on the size of training data on the corresponding client,
that is, the larger the size of the training data, the greater the contribution of the client to the global model.
This is reasonable and natural in synchronous FL.
However, in this paper,
in order to achieve good round efficiency while making full use of data from heterogeneous devices,
we use a semi-asynchronous model update and staleness tolerant distribution scheme.
At round $r$, the version of the model based on which the client $C_i$ performs local training may not be up-to-date,
namely $r_i$ is not necessarily equal to $r$.
Meanwhile, the performance of FL depends on the quality of local models.
Greater staleness may lead to larger errors when aggregating the global model.
Therefore, it is not reasonable that all participating local models are weighted solely based on the data size regardless of their model versions,
in other words, local clients based on different model versions should not be considered equally important.

In order to make up for the above shortcoming, we further optimize the aggregation function, which takes the staleness of the local models into account,
that is, the greater the staleness of the local model, the lower its weight on the aggregation should be.
The aggregation function can be expressed as
\begin{equation}
  \omega_g^{r+1} = f(r) * \omega_s^{r+1} + (1-f(r)) * \sum_{i=1}^{C*M} \frac{|D_i|}{|D_c|} * g(r-r_i) * \omega_i^{r_i+1} \label{Aggregation}
\end{equation}
where $g(r-r_i)$ is the staleness function.
In general, $g(r-r_i)$ should be $1$ when $r_i = r$ and monotonically decreases with the increase of $r-r_i$.
There are many functions that satisfy such two properties, with different rates of decline,
e.g., $g(r-r_i) = {a}^{-(r-r_i)}, a > 1$.

\subsubsection{Group-based Aggregation Function}
The non-IID nature of data leads to the model diversity among different clients,
which brings significant difficulties to the training and aggregation \cite{Group-Based}.
To address this issue, we propose a group-based aggregation function for the robust optimization and convergence of the model.
We utilize the K-means clustering method to divide clients into $|G|$ groups based on the data distribution of clients,
and then perform the global aggregation.
Specifically, after clients complete the FSSL training and upload their local models in a semi-asynchronous manner,
the server divide them into $|G|$ groups, and then updates the global model according to the following formula.
\begin{equation}
  \left\{\begin{array}{l}
    \omega_g^{r+1} = f(r) * \omega_s^{r+1} + (1-f(r)) * \frac{\sum_{k=1}^{|G|} \omega_{G_k}^{r+1}}{|G|} \\
    \omega_{G_k}^{r+1} = \sum_{C_i \in G_k} \frac{|D_i|}{|D_{G_k}|} * g(r-r_i) * \omega_i^{r_i+1}
  \end{array}\right.
\end{equation}
where $|D_{G_k}| = \sum_{C_i \in G_k} |D_i|$.
The weighted average is performed within the group based on the data size, and the adaptive attenuation is implemented according to the staleness function,
while the simple arithmetic average is used between the groups to obtain the unsupervised learning parameter of clients.
Treating groups with different data distributions as equally important (i.e., assigning the same weights) is beneficial to the robustness of the trained model.


\subsection{Adaptive Learning Rate}\label{adr}
Semi-asynchronous model updates and staleness tolerant distribution bring not only the version diversity of local models,
but also the difference in the frequency of clients participating in the global aggregation due to the device heterogeneity.
We denote the relative frequency that the client $C_i$ participates in the global update as $f_i$, and obviously $\sum_{i=1}^{M} f_i = 1$.
Then, we apply adaptive learning rate $\eta_i$ for each client based on its relative frequency for better learning performance.
Intuitively, if a client participates in the global updates more frequently, due to smaller data size, larger computing capacity, or shorter communication time,
its learning rate should accordingly decrease;
otherwise, the learning rate should rise so that the model can learn more knowledge from its data.
The corresponding adaptive learning rate $\eta_i$ of the client $C_i$ can be expressed as
\begin{equation}
  \eta_i = \frac{\lambda}{M * f_i} \label{lr}
\end{equation}
where $\lambda$ is the global learning rate, $M$ is the number of clients, and $f_i$ is the relative frequency of the client $C_i$.

A naive approach to measure the frequency is to count the number of times each client has participated in the global update in the past rounds,
and then perform the simple weighted average directly, which can be denoted as
\begin{equation}
  f_i = \frac{n_i}{\sum_{j=1}^M n_j}
\end{equation}
where $n_i$ is the number of times that the client $c_i$ has participated in the global update.
All past rounds are regarded equally and given the same weight, which is natural but suffers from an ill-consideration.
Taking Fig. \ref{Fig.Asyn} as an example, the client $C_1$ participates the global update at the rounds $r_0$ and $r_1$,
$C_2$ participates the global update at the rounds $r_0$ and $r_2$,
and $C_3$ participates the global update at the rounds $r_1$ and $r_3$.
All three of them participate in two rounds of global updates, whose frequencies will be considered the same according to the above formula.
However, intuitively, the influence of the local model $\omega_i^{r^\prime+1}$ uploaded by the client $C_i$ at the round $r^\prime$ on the global model $\omega_g^{r+1}$ gradually decays with the increase of the current training round $r$.
In other words, the global model will gradually forget knowledge from local models uploaded in the previous rounds and be replaced by the knowledge in more recent rounds.
As a result, the more recent the participation rounds of the client and the newer the historical rounds of the client, the greater the impact on the global model.
The impacts of $C_1$, $C_2$ and $C_3$ on the global model,
in descending order, should be as follows: $C_3, C_2, C_1$.
Therefore, more recent rounds need to be given greater weights while older rounds will be given lower weights.
There are many round-weight functions which can satisfy this requirement, such as
$h(r) = (1 + a)^r, a > 0$, the one inspired by the exponential smoothing \cite{ES}.

\subsection{Communication Cost Optimization}
In federated learning for anomaly detection in IoT networks, the communication cost is mainly related to two parameters,
one is the data size of exchanged model parameters at each round, and the other is the number of rounds required for model convergence.
On the one hand, the communication cost can be reduced by increasing the number of epochs for each local training.
However, as the number of epochs increases, the diversity between different local models also increase, especially on non-IID data.
On the other hand, model compression and reconstruction techniques such as structured updates and sketched updates \cite{Communication} are exploited to reduce the exchanged data size per round.
However, many compression methods also inevitably exacerbate the error of local model parameters and the diversity between different models.
The aggregation of these diverse models may result in a decrease in the accuracy of model performance and speed of model convergence.

Therefore, in order to further reduce the communication cost on the premise of ensuring model quality and convergence speed,
in this paper, we design a simple but effective approach.
We additionally add L1-regularization on the model parameter $\omega$ so that it is sparse.
First, after completing the local training, the client $C_i$ will subtract the updated model parameter from the global parameter based on which the client performs training to obtain the learned knowledge at this round of training,
such that $\Delta \omega = \omega_{i}^{r_i+1}-\omega_{g}^{r_i}$.
Then the difference between them will be transmitted as a sparse matrix to reduce communication cost.
Meanwhile, after the global update, the server also only sends the sparse matrix of differences to each client that needs to be updated (i.e., latest and deprecated clients).

\subsection{FedS$^3$A}
After describing the main components of the FedS$^3$A mechanism, we summarize its workflow as follows.
Meanwhile, the pseudo-code of our proposed FedS$^3$A mechanism is shown in Algorithm \ref{alg:1}.
\begin{enumerate}
  \item Model Initialization: At the beginning of the training (i.e., the first round $r_0$), the server $S$ randomly initializes the model parameter,
        based on which it uses its local labeled data $D_s$ to perform supervised learning for a certain number of rounds $E_s$.
        Then, it distributes the updated model parameter $\omega_g^0$ to each client participating in the federated learning $C_i, 1 \leq i \leq M$.
  \item Local Unsupervised Learning: At the current round $r$, each client $C_i$ uses the local unlabeled data samples $D_i$ combined with the pseudo-label technology to perform $E$ epochs of localized self-learning training based on the received global parameter $\omega_g^{r_i}$.
        Then, it uploads the updated local model parameter $\omega_i^{r_i+1}$ to the server once the training is completed.
        It is worth noting that due to the differences in the data size, computing capacity, communication bandwidth, etc., or due to the network failure and system crash,
        the required time for different clients to finish the training varies greatly, leading to a skew between local model versions.
  \item Global Aggregation: While the server performs supervised learning, it asynchronously receives model parameters from clients.
        After receiving the parameters from any $C$ proportion of clients, the server use Eq. (\ref{Aggregation}) to implement global aggregation without waiting for all clients to upload their model parameters.
  \item Model Distribution: After the global aggregation, the server distributes the updated global model parameter $\omega_g^{r+1}$ to the clients that need to be updated based on the version information, namely latest and deprecated clients.
  \item Repeat Step 2-4 until the global model converges or the training round reaches the specified maximum number of rounds $R$.
\end{enumerate}

\begin{algorithm}[htb]
  \caption{FedS$^3$A: Federated Semi-Supervised and Semi-Asynchronous learning for anomaly detection in IoT networks}
  \label{alg:1}
  \begin{algorithmic}[1] 
    \Require
    Maximum number of rounds $R$, Client $C_i, 1 \leq i \leq M$, Local mini-batch size $B$, number of local epochs $E$, learning rate $\eta$, staleness tolerance $\tau$
    \Ensure
    Global model $\omega_g^{R+1}$
    \State $r = 0$
    \State \textbf{Server process:}
    \For{the current round $r \leq R$}
    \If{r = 0}
    \State \Call{Initialize}{$\omega_s^0$}
    \State $\omega_g^{0}$ = \Call{LocalTraining}{$S, 0, E_s, D_s$}
    \Else
    \State $\omega_s^{r} = \omega_g^r$
    \State $\omega_s^{r+1}$ = \Call{LocalTraining}{$S, r, E, D_s$}
    \EndIf
    \EndFor
    \State $V = \emptyset$
    \While{$|V| < C*M$}
    \State Receive the local model parameter $\omega_i^{r_i+1}$ from the client $C_i$
    \State $V = V \cup \{C_i\}$
    \EndWhile
    \State Update the global model $\omega_g^{r+1}$ by Eq. (\ref{Aggregation})
    \For{each $C_i, 1 \leq i \leq M$}
    \If{$C_i \in V$ or $r - r_i > \tau$}
    \State Calculate the corresponding adaptive learning rate $\eta_i^{r+1}$ for the client $C_i$ by Eq. (\ref{lr})
    \State Distribute the current global model $\omega_g^{r+1}$ and $\eta_i^{r+1}$ to the client $C_i$
    \EndIf
    \EndFor

    \State

    \State \textbf{Client process:}
    \If{the client $C_i$ receives the global model parameter $\omega_g^r$ from the server $S$}
    \State  $\omega_i^r = \omega_g^r$
    \State $\omega_i^{r_i+1}$ = \Call{LocalTraining}{$C_i, r, E, D_i$}
    \State Upload the updated local model parameter $\omega_i^{r_i+1}$ to the server $S$
    \EndIf

    \State

    \Function{LocalTraining}{$C_i, r, E, D_i$}
    \For{epoch $e = 1$ to $E$}
    \For{mini-batch $B \in D_i$}
    \State $\omega_i^{r+1} = \omega_i^r - \eta_i^r \nabla F_i(\omega_i^r)$
    \EndFor
    \EndFor
    \State \textbf{return} $\omega_i^{r+1}$
    \EndFunction

  \end{algorithmic}
\end{algorithm}

\section{Experimental Results}\label{Experiment}
In this section, we comprehensively evaluate the performance of our FedS$^3$A scheme on anomaly detection in IoT networks.

\begin{table*}[htbp]
  \caption{Description of the basic and balanced scenarios}
  \begin{center}
    \scalebox{0.79}{
      \begin{tabular}{cccccccccccccc}
        \toprule
        \textbf{Scenario}                 & \textbf{Party} & \textbf{Total Samples} & \textbf{Benign} & \textbf{DoS Hulk} & \textbf{PortScan} & \textbf{DDoS} & \textbf{DoS GoldenEye} & \textbf{FTP-Patator} & \textbf{SSH-Patator} & \textbf{DoS slowloris} & \textbf{DoS Slowhttp} & \textbf{Entropy} \\
        \midrule
        \multirow{10}*{Basic Scenario}    & 0              & 78357                  & 4184            & 37744             & 19774             & 12784         & 1224                   & 884                  & 562                  & 524                    & 677                   & 0.5981           \\
                                          & 1              & 70470                  & 64408           & 16                & 0                 & 0             & 0                      & 1189                 & 1674                 & 1551                   & 1632                  & 0.1794           \\
                                          & 2              & 66164                  & 10592           & 19480             & 34056             & 1044          & 992                    & 0                    & 0                    & 0                      & 0                     & 0.4880           \\
                                          & 3              & 58131                  & 52248           & 5883              & 0                 & 0             & 0                      & 0                    & 0                    & 0                      & 0                     & 0.1423           \\
                                          & 4              & 44800                  & 256             & 22000             & 16072             & 5456          & 1016                   & 0                    & 0                    & 0                      & 0                     & 0.4729           \\
                                          & 5              & 39193                  & 960             & 18728             & 8517              & 10724         & 264                    & 0                    & 0                    & 0                      & 0                     & 0.5054           \\
                                          & 6              & 31211                  & 549             & 19696             & 9368              & 0             & 588                    & 0                    & 0                    & 478                    & 532                   & 0.4043           \\
                                          & 7              & 24740                  & 24740           & 0                 & 0                 & 0             & 0                      & 0                    & 0                    & 0                      & 0                     & 0                \\
                                          & 8              & 23034                  & 1008            & 8764              & 0                 & 8764          & 1788                   & 1855                 & 855                  & 0                      & 0                     & 0.6062           \\
                                          & 9              & 16904                  & 776             & 8064              & 8064              & 0             & 0                      & 0                    & 0                    & 0                      & 0                     & 0.3681           \\
        \midrule
        \multirow{10}*{Balanced Scenario} & 0              & 78356                  & 26848           & 23744             & 16465             & 7308          & 1322                   & 800                  & 665                  & 579                    & 625                   & 0.6553           \\
                                          & 1              & 70470                  & 24146           & 21354             & 14808             & 6573          & 1189                   & 719                  & 598                  & 521                    & 562                   & 0.6553           \\
                                          & 2              & 66163                  & 22670           & 20049             & 13903             & 6171          & 1116                   & 675                  & 562                  & 489                    & 528                   & 0.6553           \\
                                          & 3              & 58132                  & 19918           & 17615             & 12215             & 5422          & 981                    & 593                  & 494                  & 430                    & 464                   & 0.6553           \\
                                          & 4              & 44800                  & 15350           & 13576             & 9414              & 4179          & 756                    & 457                  & 380                  & 331                    & 357                   & 0.6553           \\
                                          & 5              & 39195                  & 13429           & 11877             & 8236              & 3656          & 661                    & 400                  & 333                  & 290                    & 313                   & 0.6553           \\
                                          & 6              & 31211                  & 10694           & 9458              & 6558              & 2911          & 527                    & 318                  & 265                  & 231                    & 249                   & 0.6553           \\
                                          & 7              & 24740                  & 8477            & 7497              & 5199              & 2308          & 417                    & 252                  & 210                  & 183                    & 197                   & 0.6553           \\
                                          & 8              & 23034                  & 7892            & 6980              & 4840              & 2148          & 389                    & 235                  & 196                  & 170                    & 184                   & 0.6553           \\
                                          & 9              & 16904                  & 5792            & 5122              & 3552              & 1577          & 285                    & 172                  & 144                  & 125                    & 135                   & 0.6553           \\

        \bottomrule
      \end{tabular}}
    \label{dataset}
  \end{center}
\end{table*}

\subsection{Dataset and Scenarios}
In this paper, we use CIC-IDS 2017 \cite{CIC-IDS} dataset for experimental evaluation,
as it is one of the largest datasets in the anomaly detection field. It can well reflect the real network environment, and has been widely used.
The dataset contains 78-dimension attribute features and 1-dimension classification feature which contains 15 attack types.
We discard several attacks with very sparse data, and the final experimental dataset contains 8 attack types: DoS Hulk, PortScan, DDoS, DoS GoldenEye, FTP-Patator, SSH-Patator, DoS slowloris, and DoS Slowhttp.
The whole dataset has about 3 million data, however, due to the extremely unbalanced distribution of categorizes,
that is about 75 percentage of the data is benign,
we select about 540,000 data for experiment, of which the ratio of training and test-validation data is about 9:1.
Meanwhile, the labeled training data of the server accounts for about 5 percentage of the total training data.

Furthermore, in order to evaluate the performance of our scheme under different data distributions, we design two different scenarios, namely basic and balanced scenarios, whose data distributions are shown in Table \ref{dataset}.
In the basic scenario, the distribution of classes and samples among different clients is highly unbalanced.
It represents a real IoT network in which some devices work normally and perform the expected operations while others suffer from multiple attacks.
In the balanced scenario, the data of each client is independent and identically distributed, and only the amount of their data is different.
Meanwhile, Shannon Entropy \cite{Shannon} is used to measure the imbalance of the local data of each client $C_i$,
which can be given by
\begin{equation}
  Entropy_i = \frac{-\sum_{k=1}^{K} \frac{|D_{i_k}|}{|D_i|} \log \frac{|D_{i_k}|}{|D_i|}}{\log K}
\end{equation}
where $K$ is the data classes of the client $C_i$ where the number of each class is $|D_{i_k}|$.
It is equal to 0 when the number of all classes is 0 expect one class, and 1 when the number of each class is equal.

\subsection{Experiment Setup}
We use two mainstream machine learning frameworks, Keras Library and TensorFlow backend \cite{Keras} to implement our experiments,
that employs a federated architecture for learning.
The system is set up with a security service provider (i.e., the server) and 10 security gateways (i.e., clients), each of which deploys an anomaly detection component and monitors several heterogeneous IoT devices.
The CNN model is used for anomaly detection which consists of two 1D-CNN (the filters are 128 and 256 respectively), one flatten,
one fully-connected (the number of hidden neurons is 256 and the activation function is ReLU), one dropout (the dropout rate is 0.1)
and one fully-connected (the activation function is Softmax) layers.
Table \ref{parameter} summarizes other default experimental settings.

The federated learning environment is simulated in a Lenovo workstation with 80-core Intel(R) Xeon(R) Gold 6230N CPU @ 2.30GHz, and 352 GB of RAM.
Although in the simulations, federated learning is all performed on a single physical machine, it is split into several independent processes or threads,
each one acting as a client or server to perform federated-learning training in parallel.
%




\begin{table}[htbp]
  \caption{Default Simulation Settings}
  \begin{center}
    \scalebox{0.9}{
      \begin{tabular}{ccc}
        \toprule
        \textbf{}                                    & \textit{Basic}             & \textit{Balanced}       \\
        \midrule
        \textbf{Staleness Function}                  & {Exponential}              & {Polynomial}            \\

        \textbf{Round-weight Function}               & {Exponential}              & {Exponential Smoothing} \\
        \textbf{Staleness Tolerance}                 & \multicolumn{2}{c}{2}                                \\
        \textbf{Participation Proportion of Clients} & \multicolumn{2}{c}{0.6}                              \\
        \textbf{Optimizer}                           & \multicolumn{2}{c}{Adam}                             \\
        \textbf{Learning Rate}                       & \multicolumn{2}{c}{0.0001}                           \\
        \textbf{Batch Size}                          & \multicolumn{2}{c}{100}                              \\
        \textbf{Epoch}                               & \multicolumn{2}{c}{1}                                \\
        \textbf{Pseudo-labeling Threshold}           & \multicolumn{2}{c}{0.95}                             \\
        \bottomrule
      \end{tabular}
    }
    \label{parameter}
  \end{center}
\end{table}

\subsection{Evaluation Metrics}
In this paper, in order to comprehensively analyze the experimental results from different perspectives,
we consider the following metrics:
\begin{enumerate}
  \item $Accuracy = \frac{TP+TN}{TP+FP+FN+TN}$
  \item $Precision = \frac{TP}{TP+FP}$
  \item $Recall/TPR = \frac{TP}{TP+FN}$
  \item $F1-Score = \frac{2TP}{2TP+FN+FP}$
  \item $FPR = \frac{FP}{FP+TN}$
\end{enumerate}
where $TP$ represents true positives, $TN$ represents true negatives, $FP$ represents false positives, and $FN$ represents false negatives.
Since CIC-IDS 2017 has nine output categorizes, which is different from traditional binary classification,
and the basic scenario is based on the imbalanced dataset,
in this paper,
we use a weighted average method to calculate these metrics.
First, we compute these metrics for each class independently,
and then a weighted average is performed based on the number of each class to obtain the final result.

At the same time, in order to conduct experimental analysis on round efficiency and communication cost, in some experiments, we further use the following metrics:
\begin{enumerate}
  \setcounter{enumi}{5}
  \item Average communication overhead (ACO): The average ratio of the data communicated between clients and the server to the total model parameters per round.
  \item Average round time (ART): The average time from distributing the global model to implementing the model update at each round.
\end{enumerate}

In addition, to avoid bias, each experiment is simulated for 10 rounds and the average value is used as the final experimental result.

\subsection{Parameter Selection}
\subsubsection{The impact of different staleness functions}\label{stalenessfunctions}
In this section, we evaluate the impact of different staleness functions introduced in  Section \ref{staleness} on federated learning performance,
and we design the following four different functions that satisfy the conditions:
\begin{itemize}
  \item Constant: $g_1(r-r_i) = 1$
  \item Polynomial: $g_2(r-r_i) = {(r-r_i+1)}^{-a}$
  \item Hinge: $g_3(r-r_i) =   \begin{cases}
            1,                          & {\text{if}}\  r-r_i \leq b \\
            {\frac{1}{a(r-r_i+b) + 1},} & {\text{otherwise.}}
          \end{cases}$
  \item Exponential: $g_4(r-r_i) = {a}^{-(r-r_i)}$
\end{itemize}

We conduct extensive experiments, and due to space limitations, we only show some key experimental results.
The results summary is shown in Table \ref{stalefunction}.
For the polynomial function $g_2$, $a = \frac{1}{2}$;
for the hinge function $g_3$, we take $a = 1, b = 0$;
for the exponential function $g_4$, $a = \frac{e}{2}$.

The constant function (i.e., $g_1$) means no adaptive decay for the weights of the staleness of local models.
It is obvious that in both experimental scenarios, each other staleness function (i.e., $g_2 \sim g_4$) can effectively improve the performance of the trained model.
This is because FedS$^3$A use a semi-asynchronous model update and staleness tolerant distribution scheme, and
taking the staleness of local models into consideration can effectively alleviate the poisoning of outdated local models to the global model,
while making full use of the data of each client.
It can be seen that in the basic scenario, the exponential function improves the performance of the model the most,
and in the balanced scenario, the polynomial function improves the most.
Therefore, in the following experiments, we separately use these two staleness functions as default settings.

\begin{table}[htbp]
  \caption{Results Summary of Different Staleness Functions}
  \begin{center}
    \scalebox{0.8}{
      \begin{tabular}{ccccccc}
        \toprule
        \textbf{}               & \textbf{}   & \textbf{Accuracy} & \textbf{Precision} & \textbf{Recall/TPR} & \textbf{F1-Score} & \textbf{FPR}    \\
        \midrule
        \multirow{4}*{Basic}    & Constant    & 0.9708            & 0.9708             & 0.8489              & 0.8983            & 0.0071          \\
                                & Polynomial  & 0.9780            & 0.9780             & 0.8586              & 0.9071            & \textbf{0.0022} \\
                                & Hinge       & 0.9785            & 0.9785             & 0.8597              & 0.9065            & 0.0063          \\
                                & Exponential & \textbf{0.9818}   & \textbf{0.9818}    & \textbf{0.8965}     & \textbf{0.9312}   & 0.0059          \\
        \midrule
        \multirow{4}*{Balanced} & Constant    & 0.9688            & 0.9687             & 0.8241              & 0.8799            & 0.0083          \\
                                & Polynomial  & \textbf{0.9815}   & \textbf{0.9815}    & \textbf{0.8684}     & \textbf{0.913}    & \textbf{0.0032} \\
                                & Hinge       & 0.9781            & 0.9781             & 0.8424              & 0.893             & 0.0082          \\
                                & Exponential & 0.9810            & 0.9810             & 0.8669              & 0.9119            & 0.0034          \\
        \bottomrule
      \end{tabular}}
    \label{stalefunction}
  \end{center}
\end{table}

\subsubsection{The impact of different round-weight functions}\label{roundweightfunctions}
In Section \ref{adr}, we introduce the adaptive learning rate,
where the client adaptively adjusts its learning rate based on the frequency and number of times it participates in the global updates.
Meanwhile, as the round increases, the corresponding weight of this round should also increase.
Therefore, in this section, we explore the impact of adaptive learning rate and different round-weight functions on the performance as follows.
\begin{itemize}
  \item Constant: $h_1(r) = 1$
  \item Logarithmic: $h_2(r) = \ln (1+r)$
  \item Polynomial: $h_3(r) = (1+r)^a$
  \item Exponential Smoothing: $h_4(r) = (1+a)^r$
  \item Exponential: $h_5(r) = a^r$
\end{itemize}

Due to the space limitations, we present some key experimental results.
For the polynomial function $h_3(r)$, $a = \frac{1}{2}$;
for the exponential smoothing $h_4(r)$, $a = 0.1$;
for the exponential function $h_5(r)$, $a = \frac{e}{2}$.

As shown in Table \ref{frefunction}, in the two experimental scenarios,
the other four functions (i.e., $h_2 \sim h_5$) have different degrees of improvements in the model performance
compared with the simple weighted average (i.e., the constant function $h_1$).
The reason is that assigning higher weights to more recent rounds can better balance how well the global model learns from the data of each client,
thereby improving the model performance.
Exponential function achieves the best performance in the basic scenario, while the exponential smoothing performs best in the balanced scenario.
Therefore, in the following experiments, exponential function and exponential smoothing are used as the default settings for the round-weight function in the basic and balanced scenarios respectively.

Furthermore, we conduct the ablation experiment with the adaptive learning rate.
The experimental results show that the performances of the models with adaptive learning rate using five round-weight functions are all better than that of the model without adaptive learning rate.
This is because due to the heterogeneity of devices, the frequency and number of times that each client participates in the global updates is also different.
Giving all clients the same learning rate at this point makes the trained model more biased towards the knowledge learned from those clients that participate more frequently,
resulting in the worst model performance.

\begin{table}[htbp]
  \caption{Results Summary of Different Round-Weight Functions}
  \begin{center}
    \scalebox{0.7}{
      \begin{tabular}{ccccccc}
        \toprule
        \textbf{}               & \textbf{}             & \textbf{Accuracy} & \textbf{Precision} & \textbf{Recall/TPR} & \textbf{F1-Score} & \textbf{FPR}    \\
        \midrule
        \multirow{6}*{Basic}    & Non-Adaptive          & 0.9678            & 0.9678             & 0.8404              & 0.8909            & 0.0090          \\
                                & Constant              & 0.9698            & 0.9698             & 0.8444              & 0.8952            & 0.0068          \\
                                & Logarithmic           & 0.9712            & 0.9712             & 0.8545              & 0.9023            & 0.0051          \\
                                & Polynomial            & 0.9739            & 0.9739             & 0.8599              & 0.9076            & 0.0027          \\
                                & Exponential Smoothing & 0.9737            & 0.9737             & 0.8578              & 0.9066            & \textbf{0.0021} \\
                                & Exponential           & \textbf{0.9818}   & \textbf{0.9818}    & \textbf{0.8965}     & \textbf{0.9312}   & 0.0059          \\
        \midrule
        \multirow{6}*{Balanced} & Non-Adaptive          & 0.9568            & 0.9568             & 0.8028              & 0.8630            & 0.0104          \\
                                & Constant              & 0.9592            & 0.9592             & 0.8088              & 0.8683            & 0.0094          \\
                                & Logarithmic           & 0.9734            & 0.9734             & 0.8574              & 0.9056            & 0.0032          \\
                                & Polynomial            & 0.9752            & 0.9752             & 0.8685              & 0.9130            & 0.0033          \\
                                & Exponential Smoothing & \textbf{0.9819}   & \textbf{0.9819}    & \textbf{0.8707}     & \textbf{0.9149}   & \textbf{0.0029} \\
                                & Exponential           & 0.9815            & 0.9815             & 0.8684              & 0.9130            & 0.0032          \\
        \bottomrule
      \end{tabular}}
    \label{frefunction}
  \end{center}
\end{table}

\subsubsection{The impact of different staleness tolerances}\label{stalenesstolerances}
Next, we test the effect of different staleness tolerances on model performance.
FedS$^3$A combines the semi-asynchronous model update and staleness tolerant distribution to achieve a trade-off between model performance and round efficiency.
For each client, if the staleness of its local model is greater than the staleness tolerance threshold,
it will be forced to update.

The experimental results are shown in Table \ref{tolerance}.
When the staleness tolerance is 0, the model performance is the worst,
because FedS$^3$A does not fully utilize the data of all clients for training.
The server starts the global aggregation once it receives 6 local models,
and other clients that do not participate in this round of global update will be forced to abort its local training and update with the new global model,
which causes that FedS$^3$A can only use the data of the 6 clients with faster training speed.

At the same time, when the staleness tolerance is greater than 2, the performance of the model is basically stable.
To further explore the reason, we find that in FedS$^3$A, the average training time per round for the client with the largest data size (i.e., $C_0$) is about 317s,
and the average training time for the client with the smallest data size (i.e., $C_9$) is about 166s.
The data size of $C_0$ is 4.64 times that of $C_9$, while the training time of $C_0$ is only 1.9 times that of $C_9$.
This means that when the staleness tolerance reaches 2, all clients can successfully participate in the federated learning process.
Therefore, in the following experiments, we select 2 as the default settings for staleness tolerance.

\begin{table}[htbp]
  \caption{Results Summary of Different Staleness Tolerances}
  \begin{center}
    \scalebox{0.9}{
      \begin{tabular}{ccccccc}
        \toprule
        \textbf{}               & \textbf{} & \textbf{Accuracy} & \textbf{Precision} & \textbf{Recall/TPR} & \textbf{F1-Score} & \textbf{FPR} \\
        \midrule
        \multirow{5}*{Basic}    & 0         & 0.9729            & 0.9729             & 0.8556              & 0.9037            & 0.0062       \\
                                & 1         & 0.9778            & 0.9778             & 0.8639              & 0.9107            & 0.0022       \\
                                & 2         & 0.9818            & 0.9818             & 0.8965              & 0.9312            & 0.0059       \\
                                & 3         & 0.9819            & 0.9819             & 0.8975              & 0.9309            & 0.0042       \\
                                & 4         & 0.9818            & 0.9818             & 0.8991              & 0.9391            & 0.0084       \\
        \midrule
        \multirow{5}*{Balanced} & 0         & 0.9733            & 0.9733             & 0.8624              & 0.9083            & 0.0042       \\
                                & 1         & 0.9786            & 0.9786             & 0.8256              & 0.8805            & 0.0084       \\
                                & 2         & 0.9819            & 0.9819             & 0.8707              & 0.9149            & 0.0029       \\
                                & 3         & 0.982             & 0.982              & 0.8725              & 0.9159            & 0.0032       \\
                                & 4         & 0.982             & 0.982              & 0.8736              & 0.9137            & 0.0081       \\

        \bottomrule
      \end{tabular}}
    \label{tolerance}
  \end{center}
\end{table}

\subsubsection{The impact of different participation proportions of clients}\label{participationproportions}
In this section, we explore the impact of different proportions of clients participating in the global aggregation per round on the model performance.
We set the participation proportion of clients to 0.1, 0.4, 0.5, 0.6 and 1,
where 0.1 represents the asynchronous federated learning and 1 illustrates synchronous federated learning.
The experimental results are shown in Table \ref{clientnumber}.

When the participation proportion of clients is 0.1, the server starts the global update once it receives an arbitrary local model,
which increases the staleness of the local model of each client dramatically and significantly.
Therefore, although FedS$^3$A uses an adaptive staleness function to alleviate the adverse effect of local model staleness on the global model,
the rapid increase in the staleness of the local model still inevitably affects the model performance, making it the worst performance.
Meanwhile, as the participation proportion of clients increases, so does the performance of the model,
the reason is that the server can utilize more clients and their local data for training at each round.

However, it is worth noting that the improvement in model performance comes at the expense of round efficiency.
As shown in Table \ref{clientnumber}, as the participation proportion of clients increases,
the average time for FedS$^3$A to complete each round of training also increases,
because the server needs to wait for a sufficient number of clients to complete local training before the global update.
To achieve a good trade-off between the model performance and round efficiency,
we select 0.6 as the default setting for our experiments.

\begin{table}[htbp]
  \caption{Results Summary of Different Participation Proportions of Clients}
  \begin{center}
    \scalebox{0.75}{
      \begin{tabular}{cccccccc}
        \toprule
        \textbf{}               & \textbf{} & \textbf{Accuracy} & \textbf{Precision} & \textbf{Recall/TPR} & \textbf{F1-Score} & \textbf{FPR} & \textbf{ART} \\
        \midrule
        \multirow{5}*{Basic}    & 0.1       & 0.9703            & 0.9703             & 0.8547              & 0.9016            & 0.0067       & 89.2         \\
                                & 0.4       & 0.9726            & 0.9726             & 0.8551              & 0.9039            & 0.0032       & 190.82       \\
                                & 0.5       & 0.9750            & 0.9750             & 0.8630              & 0.9101            & 0.0021       & 193.46       \\
                                & 0.6       & 0.9818            & 0.9818             & 0.8965              & 0.9312            & 0.0059       & 221.4        \\
                                & 1         & 0.9871            & 0.9871             & 0.9058              & 0.9337            & 0.0065       & 356.86       \\
        \midrule
        \multirow{5}*{Balanced} & 0.1       & 0.9688            & 0.9688             & 0.8424              & 0.8930            & 0.0082       & 116.52       \\
                                & 0.4       & 0.9735            & 0.9735             & 0.8635              & 0.9090            & 0.0041       & 179.1        \\
                                & 0.5       & 0.9760            & 0.9760             & 0.8717              & 0.9152            & 0.0034       & 187.12       \\
                                & 0.6       & 0.9819            & 0.9819             & 0.8707              & 0.9149            & 0.0029       & 220.45       \\
                                & 1         & 0.9888            & 0.9888             & 0.8691              & 0.9118            & 0.0037       & 428.71       \\

        \bottomrule
      \end{tabular}}
    \label{clientnumber}
  \end{center}
\end{table}

\subsubsection{The impact of different data sizes of the server}
In this section, we evaluate the impact of different data sizes of the server on the experimental results.
In the disjoint federated semi-supervised learning scenario of this paper,
the data on the server side is labeled data, and the data on the client side is all unlabeled data.
We set the ratio of the labeled data size on the server to the total training data size to 1\%, 2\%, 4\%, 5\% and 7\% respectively.

As shown in Table \ref{servernumber}, even if the labeled data on the server only accounts for 1\%,
FedS$^3$A can still achieve an accuracy of 85.91\%.
Meanwhile, as the size of labeled data of the server increases, the performance of the model gradually improves.
The reason is that the increase in labeled data size on the server is conducive to the guidance of the unsupervised training of the clients,
which can better help the clients to attach pseudo-labels to their local unlabeled data more accurately.
In this way, the value of a large amount of unlabeled data on the clients can be better exerted.
At the same time, it can be found that when the proportion of labeled data on the server increases from 2\% to 4\%, the performance of the model has a significant improvement.
However, when the labeled data size increases from 5\% to 7\%, the performance of the model tends to stabilize.
Therefore, 5\% is chosen as the default setting for the experiments.
\begin{table}[htbp]
  \caption{Results Summary of Different Data Sizes of the Server}
  \begin{center}
    \scalebox{0.85}{
      \begin{tabular}{ccccccc}
        \toprule
        \textbf{}               & \textbf{} & \textbf{Accuracy} & \textbf{Precision} & \textbf{Recall/TPR} & \textbf{F1-Score} & \textbf{FPR} \\
        \midrule
        \multirow{5}*{Basic}    & 1\%       & 0.8591            & 0.8591             & 0.6201              & 0.6973            & 0.0359       \\
                                & 2\%       & 0.8883            & 0.8883             & 0.6805              & 0.7479            & 0.0266       \\
                                & 4\%       & 0.9705            & 0.9705             & 0.8560              & 0.9023            & 0.0062       \\
                                & 5\%       & 0.9818            & 0.9818             & 0.8965              & 0.9312            & 0.0059       \\
                                & 7\%       & 0.9876            & 0.9876             & 0.8995              & 0.9359            & 0.0050       \\
        \midrule
        \multirow{5}*{Balanced} & 1\%       & 0.8702            & 0.8702             & 0.6434              & 0.7209            & 0.0350       \\
                                & 2\%       & 0.8806            & 0.8806             & 0.6642              & 0.7357            & 0.0301       \\
                                & 4\%       & 0.9674            & 0.9674             & 0.8454              & 0.8939            & 0.0082       \\
                                & 5\%       & 0.9819            & 0.9819             & 0.8707              & 0.9149            & 0.0029       \\
                                & 7\%       & 0.9875            & 0.9875             & 0.8776              & 0.9138            & 0.0063       \\

        \bottomrule
      \end{tabular}}
    \label{servernumber}
  \end{center}
\end{table}

\subsection{Ablation Study}
In this section, we perform the ablation study on FedS$^3$A.
Since the impact of the staleness function, adaptive learning rate (i.e., round-weight function),
semi-asynchronous model update (i.e., participation proportion of clients), and staleness tolerant distribution (i.e., staleness tolerance)
have been evaluated and discussed in Section \ref{stalenessfunctions}, \ref{roundweightfunctions}, \ref{participationproportions} and \ref{stalenesstolerances},
in this section we mainly experiment and analyze the impact of the group-based aggregation function and dynamic weight of supervised learning.

\subsubsection{The impact of the group-based aggregation function}
In this section, we evaluate and validate the effectiveness of the group-based aggregation function.
The experiment is only conducted in the basic scenario (i.e., the data of clients is non-IID).
The reason is that the function groups the clients based on the data distribution of clients,
and in the balanced scenario, i.e., when the data of different clients is independent and identically distributed, it will become a random group function.

\begin{figure}[htbp]
  \centering
  \includegraphics[width=0.4\textwidth]{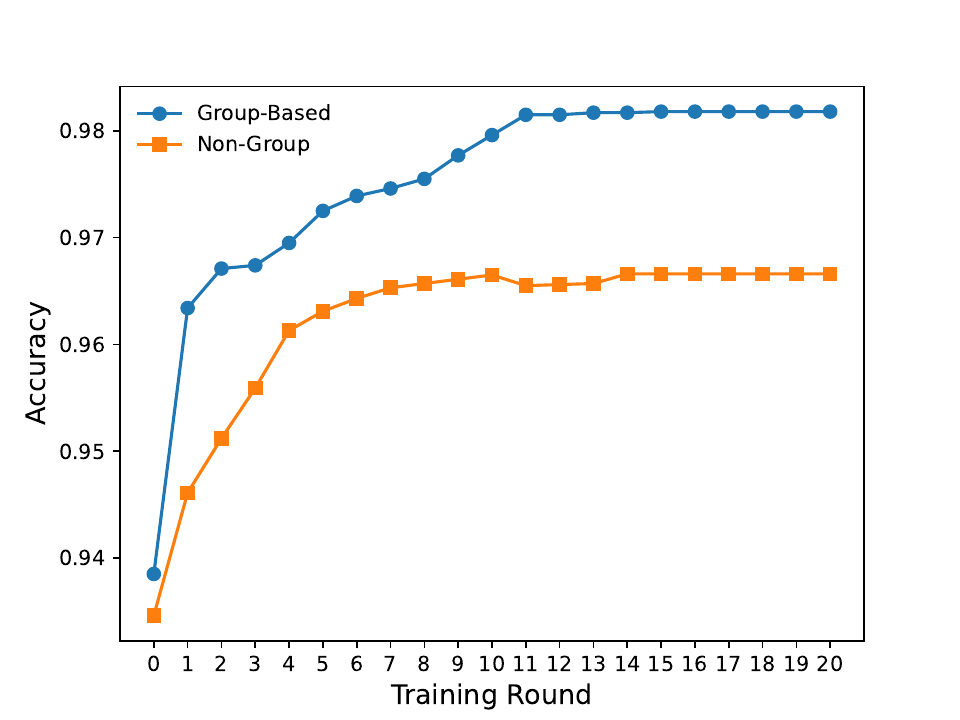}
  \caption{The Impact of Group-based Aggregation Function.}
  \label{Fig.Group}
\end{figure}

As shown in Table \ref{group} and Fig. \ref{Fig.Group}, using the group-based aggregation function can effectively improve the performance of the model.
This can be explained by the fact that FedS$^3$A can capture the main gradient features in different data distributions by grouping,
which can minimize the impact of non-IID data on the global model.

\begin{table}[htbp]
  \caption{The Impact of Group-based Aggregation Function}
  \begin{center}
    \scalebox{0.8}{
      \begin{tabular}{ccccccc}
        \toprule
        \textbf{}            & \textbf{}   & \textbf{Accuracy} & \textbf{Precision} & \textbf{Recall/TPR} & \textbf{F1-Score} & \textbf{FPR}    \\
        \midrule
        \multirow{2}*{Basic} & Non-Group   & 0.9666            & 0.9666             & 0.8357              & 0.8886            & \textbf{0.0054} \\
                             & Group-Based & \textbf{0.9818}   & \textbf{0.9818}    & \textbf{0.8965}     & \textbf{0.9312}   & 0.0059          \\
        \bottomrule
      \end{tabular}}
    \label{group}
  \end{center}
\end{table}

\subsubsection{The impact of the dynamic weight of supervised learning}
In order to verify the effectiveness of the dynamic weight of supervised learning,
we conduct a series of experiments.
As mentioned earlier, the weight of supervised learning dynamically decays from $\frac{1}{2}$ to $\frac{1}{C*M+1}$ (i.e., $\frac{1}{7}$ due to $C=0.6$ and $M=10$) as the number of rounds increases.
Meanwhile, Table \ref{aslw} also shows the experimental results when the weight of supervised learning is fixed to $\frac{1}{2}$ and $\frac{1}{7}$ respectively.

First, it is obvious that the dynamically decayed weight of supervised learning outperforms the two fixed supervised-learning weights in both basic and balanced scenarios.
This is because the dynamic attenuation of the supervised-learning weight can well solve the problem of
insufficient guidance for unsupervised learning of clients in the early stage of federated semi-supervised learning,
and can effectively avoid the problem of model overfitting due to the excessive supervised learning weight in the later stage of the training,
thereby achieving better model performance.

\begin{figure}[!t]
  \centering
  \subfloat[]{\includegraphics[width=0.48\columnwidth]{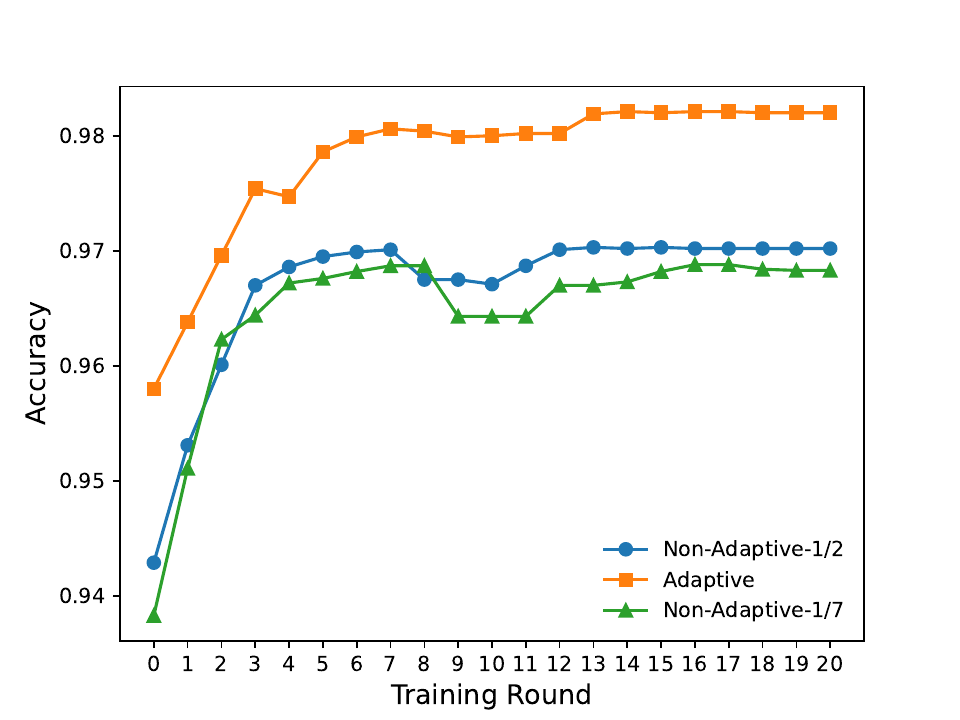}
    \label{Supervised_Weight_Basic}}
  \hfil
  \subfloat[]{\includegraphics[width=0.48\columnwidth]{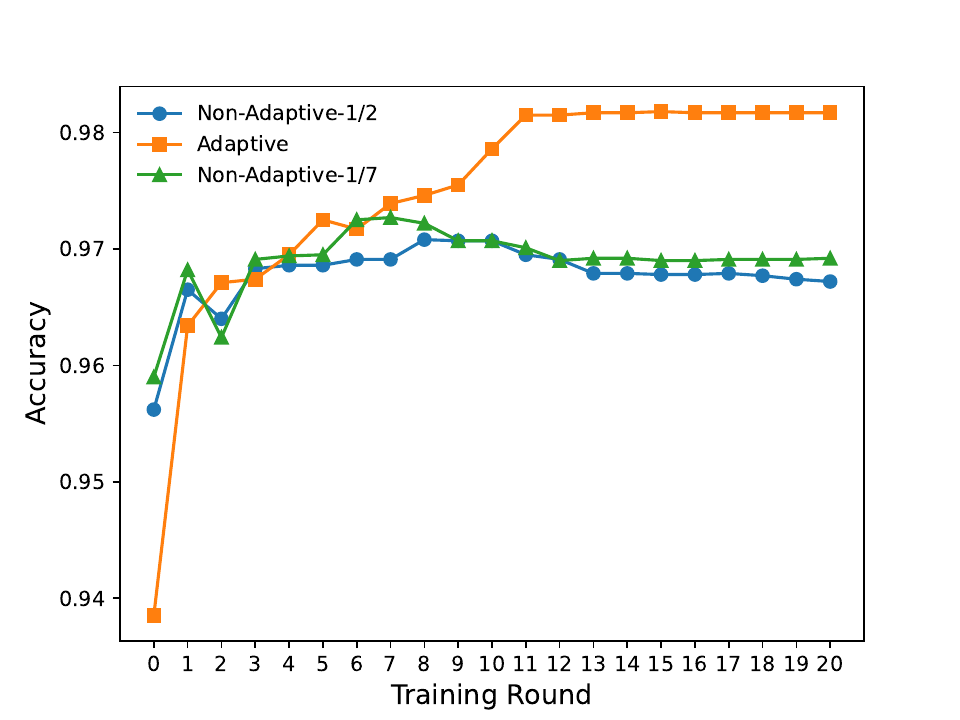}
    \label{Supervised_Weight_Balanced}}
  \caption{The impact of the dynamic weight of supervised learning on detection accuracy. (a) Basic Scenario. (b) Balanced Scenario.}
  \label{fig:Supervised_Weight}
\end{figure}

Then, as shown in Fig. \ref{fig:Supervised_Weight}, in the basic scenario, the model with a supervised-learning weight of $\frac{1}{2}$ performs better than that of $\frac{1}{7}$, while in the balanced scenario, the opposite is true.
The reason is that in the basic scenario, the data of the server and clients is non-IID.
In this case, assigning a fixed weight of $\frac{1}{7}$ to the supervised learning will cause the server can not play its guiding role in unsupervised learning on the clients.
In the balanced scenario, assigning a fixed weight of $\frac{1}{2}$ to the server will cause the trained model to rely too much on the labeled data on the server,
and can not give full play to the role of a large amount of unlabeled data on clients,
which will lead to the model overfitting.

\begin{table}[htbp]
  \caption{The Impact of Dynamic Weight of Supervised Learning}
  \begin{center}
    \scalebox{0.75}{
      \begin{tabular}{ccccccc}
        \toprule
        \textbf{}               & \textbf{}                  & \textbf{Accuracy} & \textbf{Precision} & \textbf{Recall/TPR} & \textbf{F1-Score} & \textbf{FPR}    \\
        \midrule
        \multirow{3}*{Basic}    & Non-Adaptive-$\frac{1}{2}$ & 0.9703            & 0.9703             & 0.8517              & 0.9               & 0.0055          \\
                                & Adaptive                   & \textbf{0.9818}   & \textbf{0.9818}    & \textbf{0.8965}     & \textbf{0.9312}   & \textbf{0.0059} \\
                                & Non-Adaptive-$\frac{1}{7}$ & 0.9688            & 0.9688             & 0.8471              & 0.8963            & 0.0061          \\
        \midrule
        \multirow{3}*{Balanced} & Non-Adaptive-$\frac{1}{2}$ & 0.9708            & 0.9708             & 0.8494              & 0.8991            & 0.0047          \\
                                & Adaptive                   & \textbf{0.9819}   & \textbf{0.9819}    & \textbf{0.8707}     & \textbf{0.9149}   & \textbf{0.0029} \\
                                & Non-Adaptive-$\frac{1}{7}$ & 0.9727            & 0.9727             & 0.8551              & 0.9038            & 0.0036          \\
        \bottomrule
      \end{tabular}}
    \label{aslw}
  \end{center}
  \vspace{-0.3cm}
\end{table}

\subsection{Performance Comparison}
In this section, we compare FedS$^3$A with other federated learning frameworks.
Since the algorithm design of each related work is based on their own proposed system architecture,
it is difficult to compare and analyze all works.
In this context, we summarize the other algorithms, group them into several categorizes,
and make some minor changes to allow for more objective and unbiased comparisons and analysis under the same experimental scenarios and communication frameworks, as shown below.

\subsubsection{Benchmarks}

\begin{enumerate}[label=\alph*)]
  \item FedAvg-SSL: FedAvg \cite{FedAvg} is a mainstream federated learning method and is widely used for anomaly detection in IoT networks \cite{DIoT,NBaIoTCN,ToNIoTCN,DADIoT}.
        Based on the participation proportion of clients, it can be divided into two types, FedAvg-SSL-Partial and FedAvg-SSL-All.
        In FedAvg-SSL-Partial, at each round, the server randomly pre-specify partial clients to participate in this round of FL training process,
        while in FedAvg-SSL-All, all clients should participate in each round of training.
        In this paper, for consistency, we set the number of participating clients per round for FedAvg-SSL-Partial to 6.
  \item FedAsync-SSL: FedAsync is introduced in \cite{AFL} which is a classic asynchronous federated learning algorithm,
        and its idea has been applied to anomaly detection in IoT networks \cite{AFLIoT}.
        The server starts the global update once it receives one local model from an arbitrary client.
        In this paper, we use the polynomial function as the staleness function and set $\alpha = 0.9$, $a = 0.5$, $t - \tau \leq 16$ and $\rho = 0.005$,
        which is claimed to be the combination with the best performance in \cite{AFL}.
  \item Local-SSL: At the same time, Local-SSL \cite{Fixmatch} aggregates the labeled data of the server and the unlabeled data of all clients to perform the local semi-supervised training,
        which should be the ceiling of the performance.
        It is used to evaluate and measure the gap between our distributed FedS$^3$A method and the centralized Local-SSL.
\end{enumerate}

In addition, since in the existing works, both FedAvg and FedAsync are used in the traditional federated supervised learning scenario, where there is no data on the server, and the data on the clients is all labeled.
This is different from the disjoint scenario for federated semi-supervised learning introduced in this paper.
To solve the dilemma, we add the weight of supervised learning at the server into the global update with the weights of unsupervised learning of the clients,
and adopt the dynamic weight of supervised learning with the best performance, which has been proved in the experiment.

\begin{table}[htbp]
  \caption{Results Summary of Performance Comparison}
  \begin{center}
    \scalebox{0.6}{
      \begin{tabular}{ccccccccc}
        \toprule
        \textbf{}               & \textbf{}          & \textbf{Accuracy} & \textbf{Precision} & \textbf{Recall/TPR} & \textbf{F1-Score} & \textbf{FPR}    & \textbf{ART} & \textbf{ACO}  \\
        \midrule
        \multirow{4}*{Basic}    & FedS$^3$A             & \textbf{0.9818}   & \textbf{0.9818}    & \textbf{0.8965}     & \textbf{0.9312}   & \textbf{0.0059} & 221.4        & \textbf{0.49} \\
                                & FedAvg-SSL-Partial & 0.9471            & 0.9471             & 0.7987              & 0.8530            & 0.0176          & 338.6        & 1             \\
                                & FedAvg-SSL-All     & 0.9636            & 0.9636             & 0.8382              & 0.8871            & 0.0101          & 341.65       & 1             \\
                                & FedAsync-SSL       & 0.8667            & 0.8939             & 0.6594              & 0.7510            & 0.0201          & 85.4         & 1             \\
        \midrule
        \multirow{4}*{Balanced} & FedS$^3$A             & \textbf{0.9819}   & \textbf{0.9819}    & \textbf{0.8707}     & \textbf{0.9149}   & \textbf{0.0029} & 220.45       & \textbf{0.48} \\
                                & FedAvg-SSL-Partial & 0.9641            & 0.9641             & 0.8279              & 0.8823            & 0.0070          & 321.8        & 1             \\
                                & FedAvg-SSL-All     & 0.9671            & 0.9671             & 0.8373              & 0.8897            & 0.0061          & 324.7        & 1             \\
                                & FedAsync-SSL       & 0.9599            & 0.9599             & 0.8094              & 0.8707            & 0.0056          & 88.25        & 1             \\
        \midrule
        \textbf{}               & Local-SSL          & 0.9910            & 0.9844             & 0.9479              & 0.9511            & 0.0016          & -            & -             \\
        \bottomrule
      \end{tabular}}
    \label{comparison}
  \end{center}
\end{table}

\subsubsection{Comparison Results}
As shown in Table \ref{comparison}, FedS$^3$A outperforms three FL-based comparison algorithms, namely FedAvg-SSL-Partial, FedAvg-SSL-All and FedAsync-SSL in both scenarios, especially in the basic scenario.
The performance of FedS$^3$A remains stable in both scenarios, while three comparison algorithms are adversely affected by the non-IID data, especially FedAsync-SSL.
The is because FedS$^3$A groups clients based on their data distributions to learn knowledge from clients with different data distributions, which is beneficial to the robustness of the model.
In contrast, FedAsync-SSL implements the global update once an arbitrary local model is received.
Therefore, the non-IID nature of data leads to huge differences in local models between consecutive adjacent rounds,
which results in the oscillations and declines in the model performance.
Meanwhile, the performance of FedS$^3$A is close to that of Local-SSL, even in the basic scenario.
The reason is that FedS$^3$A minimizes the impact of semi-asynchrony and non-IID data by taking the staleness of local models, different participation frequencies and different data distributions of clients into consideration.

FedS$^3$A greatly improves the round efficiency through the semi-asynchronous model update and staleness tolerant distribution scheme compared to the synchronous federated learning method.
Meanwhile, the average round time of FedAvg-SSL-Partial is almost the same as that of FedAvg-SSL-All,
even though it only randomly specifics 6 clients instead of all of them per round.
This is because in synchronous federated learning, the server must wait for all specified clients to complete training before global updates,
so the average round time is determined by the slowest client.
It is not difficult to find that if 6 clients are randomly assigned from ten clients,
the probability of assigning to $C_0$ is 60\%.
Therefore, the round efficiency of FedAvg-SSL-Partial can not be much higher than that of FedAvg-SSL-All.

In addition, by calculating the difference of sparse matrices, FedS$^3$A only needs to transmit the knowledge learned at each round of training
instead of all model parameters, thus reducing the communication cost.
Experimental results show that FedS$^3$A successfully reduces the communication cost by half through this simple approach.

\section{Conclusion}\label{Conclusion}
In this paper, we proposed a Federated Semi-Supervised and Semi-Asynchronous (FedS³A) learning framework tailored for anomaly detection in IoT networks. Our approach addresses a practical scenario where labeled data is exclusively available at the server, while each client holds substantial amounts of unlabeled data. To tackle this, FedS³A leverages federated semi-supervised learning and pseudo-labeling techniques, complemented by a dynamic weighting mechanism for supervised learning. This strategy effectively balances the contributions of the server’s labeled data with the clients’ unlabeled data during model training.

To account for the device heterogeneity prevalent in IoT networks, we designed a semi-asynchronous model update strategy paired with a staleness-tolerant distribution scheme. This allows all clients to participate in the federated training process, with the server performing global updates once parameters from a predefined proportion of clients are received. To optimize the trade-off between round efficiency and client utilization, FedS³A permits a subset of clients to operate with asynchronous model versions. The impact of stale models is mitigated by reducing their weights, while an adaptive learning rate ensures balanced contributions across clients. Additionally, we incorporated a group-based aggregation function to address the challenges posed by non-IID data distributions, enhancing model robustness.

To minimize communication overhead without compromising performance, we introduced an efficient protocol that transmits only parameter differences as sparse matrices across communication rounds. This approach significantly reduces the bandwidth requirements, a critical consideration for resource-constrained IoT environments.

Extensive experimental results validate the efficacy of FedS³A, demonstrating that it significantly outperforms state-of-the-art federated learning-based anomaly detection methods. Specifically, our framework achieves detection accuracies exceeding 98\% while reducing communication costs to less than 50

\section*{Acknowledgment}
This work was supported by 
the Special Program on Industrial Foundation Re-engineering and High-quality Development of Manufacturing Industry by Ministry of Industry and Information Technology (MIIT).

\bibliographystyle{IEEEtran}
\bibliography{FL}

\begin{IEEEbiography}[{\includegraphics[width=1in,height=1.25in,clip,keepaspectratio]{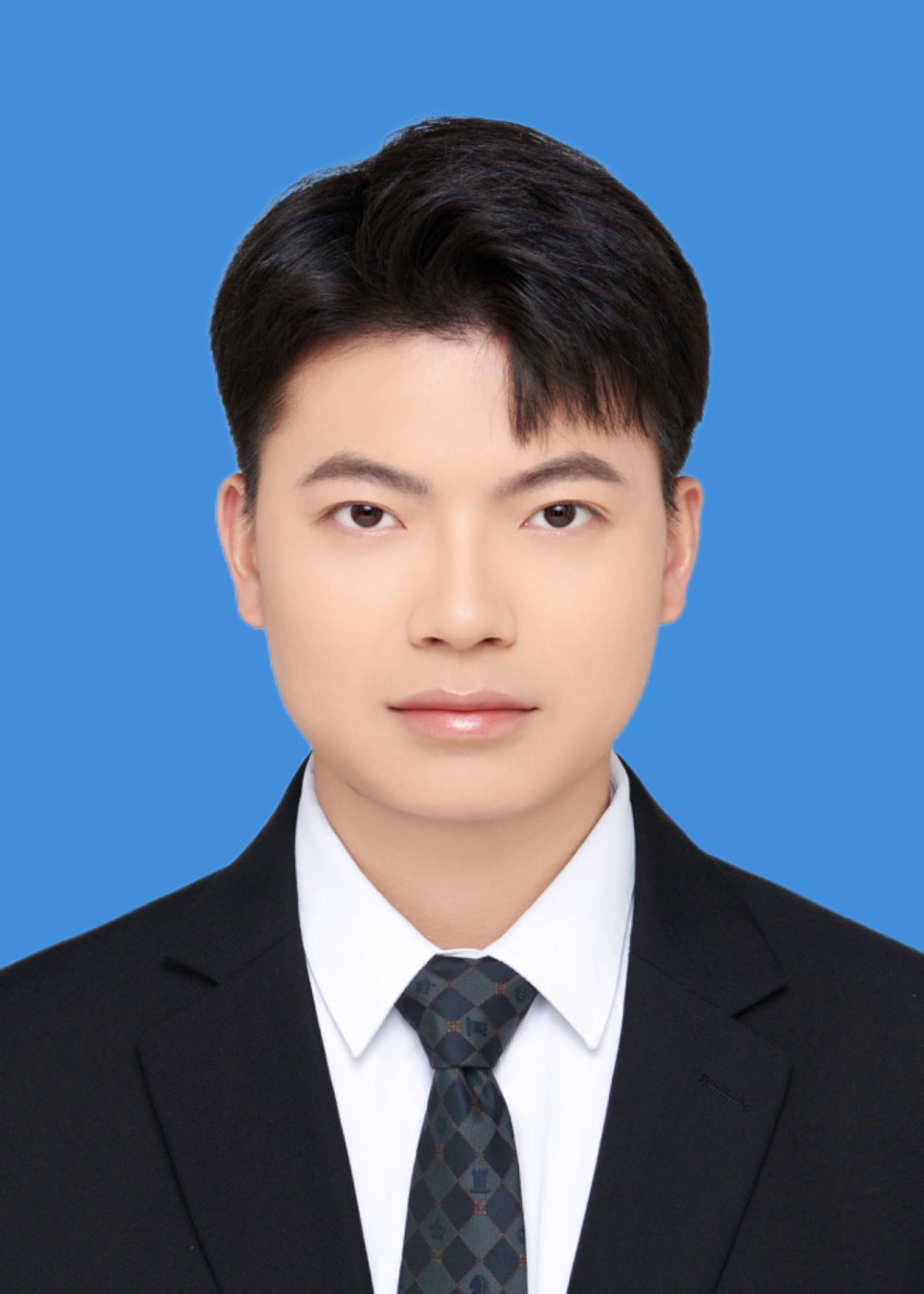}}]{Wenbin~Zhai}
  is currently pursing the Ph.D. degree with the Department of Computing, The Hong Kong Polytechnic University. He received the B.S. degree from Nanjing University of Chinese Medicine, Nanjing, Jiangsu Province, China in 2020, and the M.S. degree in Nanjing University of Aeronautics and Astronautics, Nanjing, Jiangsu Province, China in 2023. His research interests include AI security, Cybersecurity, Wireless Sensor Networks (WSNs), and Networking.
\end{IEEEbiography}

\begin{IEEEbiography}[{\includegraphics[width=1in,height=1.25in,clip,keepaspectratio]{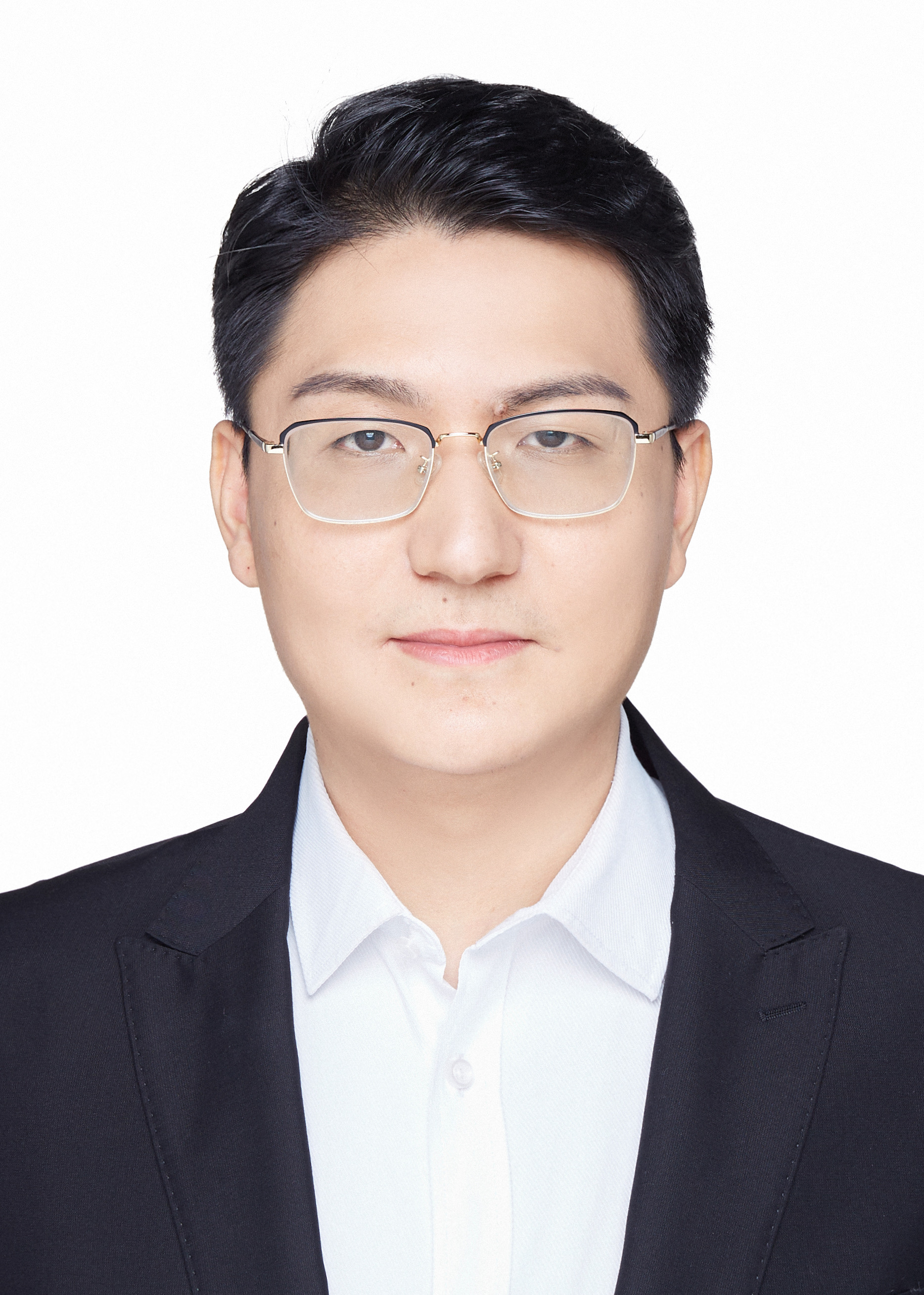}}]{Liang~Liu}
  is currently an associate professor in College of Computer Science and Technology, Nanjing University of Aeronautics and Astronautics, Nanjing, Jiangsu Province, China. His research interests include distributed system, big data and system security. He received the B.S. degree in computer science from  Northwestern Polytechnical University, Xi’an, Shanxi Province, China in 2005, and the Ph.D. degree in computer science from Nanjing University of Aeronautics and Astronautics, Nanjing, Jiangsu Province, China in 2012.
\end{IEEEbiography}

\begin{IEEEbiography}[{\includegraphics[width=1in,height=1.25in,clip,keepaspectratio]{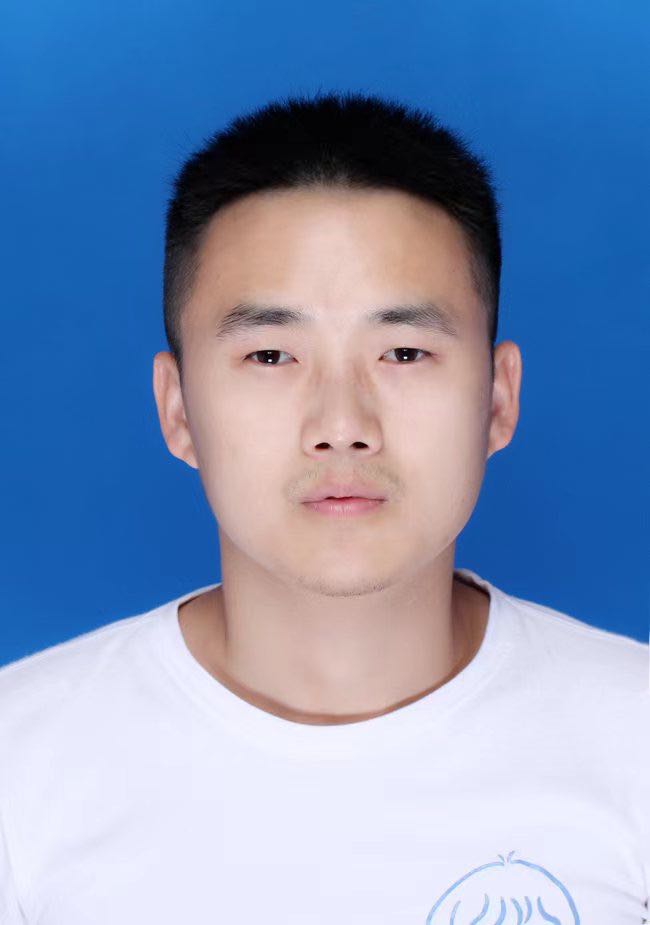}}]{Feng~Wang}
  graduated from Zhicheng College of Fuzhou University with a bachelor's degree in 2018. At present, he is studying for a master's degree in the collage of Computer Science and Technology, Nanjing University of Aeronautics and Astronautics. His main research direction is moving target defense and distributed databases.
\end{IEEEbiography}

\begin{IEEEbiography}[{\includegraphics[width=1in,height=1.25in,clip,keepaspectratio]{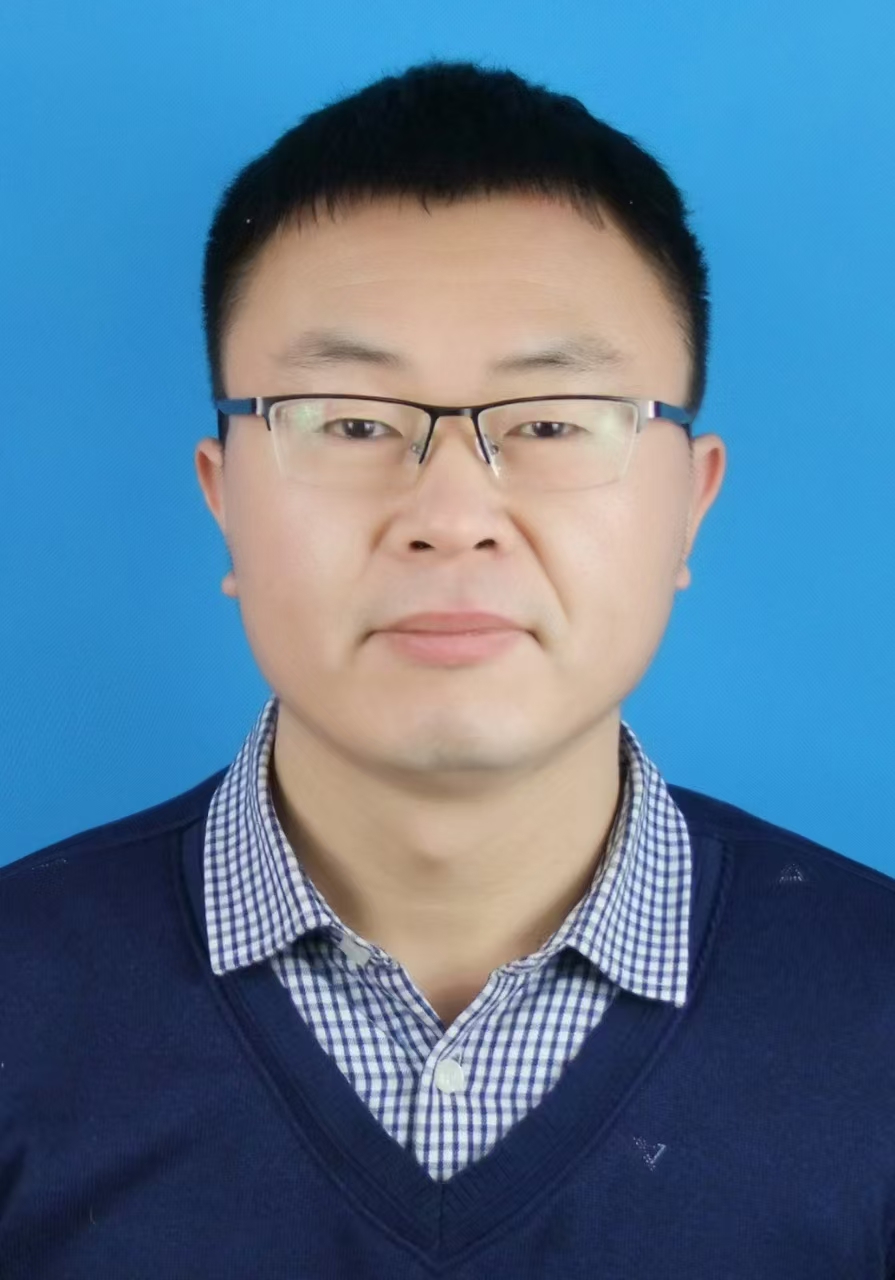}}]{Youwei~Ding}
  is currently a lecturer in the School of Artificial Intelligence and Information Technology, Nanjing University of Chinese Medicine, Nanjing, Jiangsu Province, China. His research interests include energy efficient data management, big data analysis and data security. He received the B.S and M.S degrees in computer science from Yangzhou University, Yangzhou, Jiangsu Province, China in 2007 and 2010, and the Ph.D. degree in computer science form Nanjing University of Aeronautics and Astronautics, Nanjing, Jiangsu Province, China in 2016.
\end{IEEEbiography}

\begin{IEEEbiography}[{\includegraphics[width=1in,height=1.25in,clip,keepaspectratio]{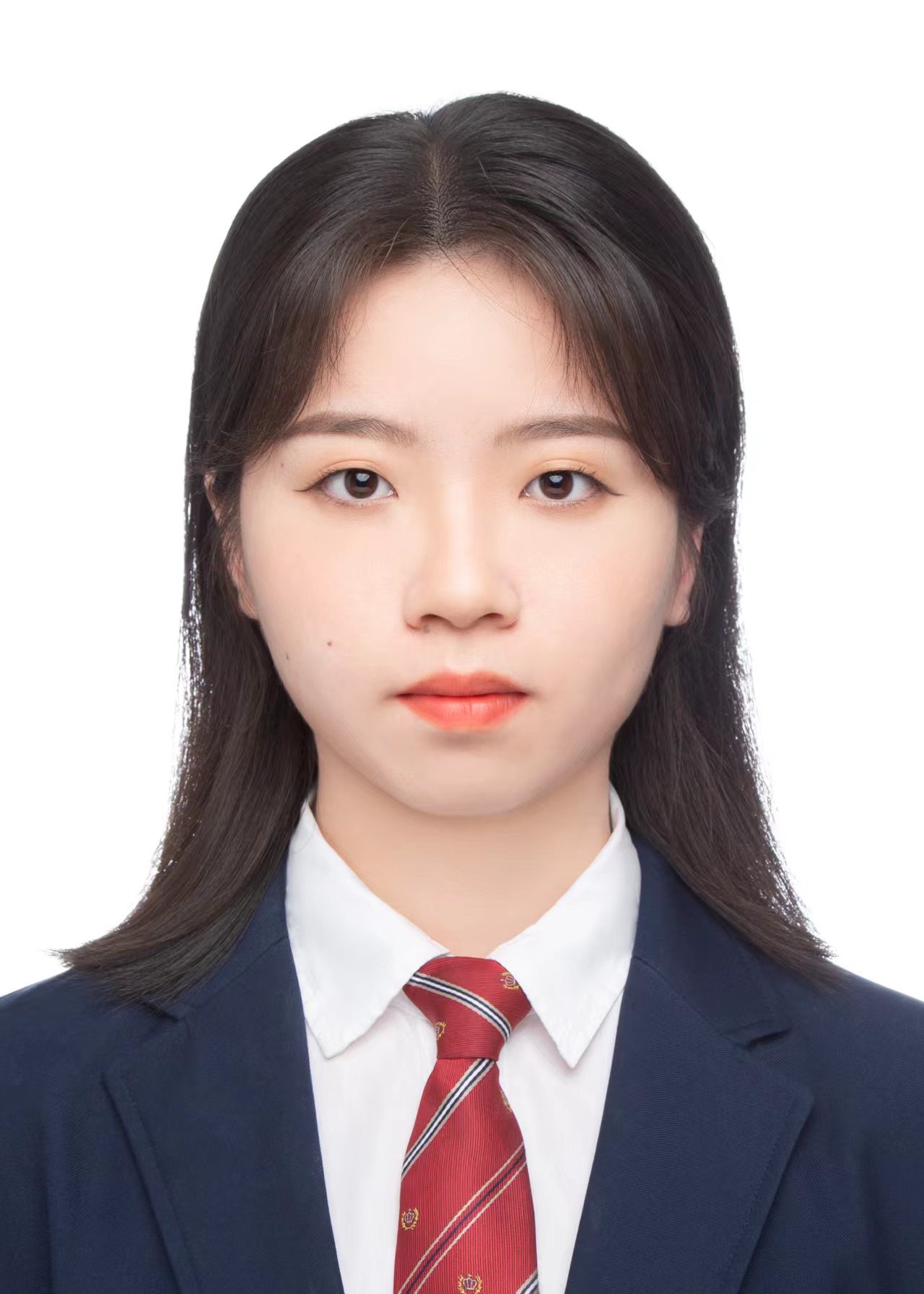}}]{Wanying~Lu}
  graduated from Henan Polytechnic University with a bachelor's degree in 2021. At present, she is studying for a master's degree in the collage of Computer Science and Technology, Nanjing University of Aeronautics and Astronautics. Her main research direction is time series big data storage and data mining.
\end{IEEEbiography}

\begin{IEEEbiography}[{\includegraphics[width=1in,height=1.25in,clip,keepaspectratio]{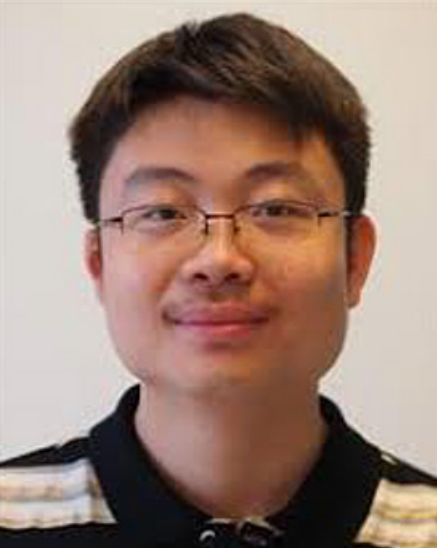}}]{Weizhi~Meng}
  received the
Ph.D. degree in computer science from the City
University of Hong Kong, Hong Kong, in 2013.
He is currently a Professor with the School
of Computing and Communications, Lancaster
University, Lancaster, U.K. He was a Faculty
Member with the Department of Applied
Mathematics and Computer Science, Technical
University of Denmark, Denmark. His primary
research interests include cyber security and
intelligent technology in security, including intrusion
detection, smartphone security, biometric authentication, HCI security, cloud
security, trust management, blockchain in security, cyber–physical system
security, and IoT security.
Prof. Meng won the Outstanding Academic Performance Award during
his doctoral study, the IEEE ComSoc Best Young Researcher Award for
Europe, Middle East, and Africa Region in 2020, and the IEEE MGA
Young Professionals Achievement Award in 2020. He is a recipient of the
Hong Kong Institution of Engineers Outstanding Paper Award for Young
Engineers/Researchers in 2014 and 2017.
\end{IEEEbiography}

\vfill

\end{document}